%% file: main.tex
\documentclass[fleqn,usenatbib]{mnras}

\usepackage{newtxtext,newtxmath}

\usepackage[T1]{fontenc}

\DeclareRobustCommand{\VAN}[3]{#2}
\let\VANthebibliography\thebibliography
\def\thebibliography{\DeclareRobustCommand{\VAN}[3]{##3}\VANthebibliography}

\usepackage{graphicx}	
\usepackage{amsmath}	

\usepackage{caption}
\usepackage{verbatim}  
\usepackage{float}
\usepackage[dvipsnames]{xcolor}     
\usepackage{siunitx}
\usepackage{rotating}
\usepackage{amsfonts}
\usepackage{hyperref}
\usepackage{cleveref}
\crefname{section}{§}{§§}
\Crefname{section}{§}{§§}
\usepackage{longtable}
\usepackage{natbib}
\usepackage{verbatim}
\usepackage{lscape}
\usepackage{graphics}
\usepackage{epsfig}
\usepackage{multirow}
\usepackage{setspace}
\usepackage{bigdelim}
\usepackage{bigstrut}
\usepackage{floatflt}	
\usepackage{wrapfig}
\usepackage{multicol}
\usepackage{indentfirst}
\usepackage{graphicx}
\usepackage{subfig}
\usepackage{lscape}

\newcommand{\angstrom}{\mbox{\ \normalfont\AA}}

\sffamily
\def\h1{H\,{\sc i}}

\def\A{\angstrom}
\def\nm{\mathrm{nm}}

\def\ga{\raisebox{-0.3ex}{\mbox{$\stackrel{>}{_\sim} \,$}}}
\def\la{\raisebox{-0.3ex}{\mbox{$\stackrel{<}{_\sim} \,$}}}

\def\c1{C\,{\sc i}}

\def\NH3{NH$_{3}$}
\def\ch3cn{CH$_{3}$CN}

\def\apj{ApJ}
\def\apjs{ApJSS}
\def\apjl{ApJL}

\def\aap{A\&A}

\def\araa{ARA\&A}

\def\aj{AJ}
\def\nat{Nature}
\def\mnras{MNRAS}

\def\slsn{SLSNe-I\ }

\def\risetime{$\tau_{\text{rise}}^{\Delta 1 \text{m}}$}

\makeatletter
\newcommand*\mysize{%
   \@setfontsize\mysize{8.0}{9.0}%
}
\makeatother

\defcitealias{inserra2014}{IS14}
\defcitealias{inserra2020}{I20}

\title[UV of SLSNe]{The restframe ultraviolet of superluminous supernovae  -- I. Potential as cosmological probes}

\author[N. Khetan et al.]{
Nandita Khetan,$^{1,2}$\thanks{E-mail: nandita.khetan@gssi.it}Jeff Cooke,$^{3,4}$ Marica Branchesi$^{2,5}$
\\
$^{1}$DARK, Niels Bohr Institute, University of Copenhagen, Jagtvej 128, DK-2200 Copenhagen \O, Denmark\\
$^{2}$Gran Sasso Science Institute, L'aquila (AQ), 67100, Italy\\
$^{3}$Centre for Astrophysics and Supercomputing, Swinburne University of Technology, Hawthorn, Victoria 3122, Australia\\
$^{4}$ARC Centre of Excellence for Gravitational Wave Discovery (OzGrav)\\
$^{5}$ INFN, Laboratori Nazionali del Gran Sasso, Assergi, Italy\\}

\date{Accepted XXX. Received YYY; in original form ZZZ}

\pubyear{2020}

\begin{document}
\label{firstpage}
\pagerange{\pageref{firstpage}--\pageref{lastpage}}
\maketitle

\begin{abstract}
Superluminous supernovae (SLSNe) have been detected to $z\sim4$ and can be detected to $z\gtrsim15$ using current and upcoming facilities. SLSNe are extremely UV luminous, and hence objects at $z\gtrsim7$ are detected exclusively via their rest-frame UV using optical and infrared facilities. SLSNe have great utility in multiple areas of stellar and galactic evolution. Here, we explore the potential use of SLSNe type-I as high-redshift cosmological distance indicators in their rest-frame UV. Using a SLSNe-I sample in the redshift range $1\lesssim z\lesssim 3$, we investigate correlations between the peak absolute magnitude in a synthetic UV filter centered at \SI{250}{\nm} and rise time, colour and decline rate of SLSNe-I light curves. We observe a linear correlation between $M_0(250)$ and the rise time with an intrinsic scatter of 0.29. Interestingly, this correlation is further tightened ($\sigma_{int} \approx 0.2$) by eliminating those SLSNe which show a pre-peak bump in their light curve. This result hints at the possibility that the ``bumpy'' SLSNe could belong to a different population. Weak correlations are observed between the peak luminosity and colour indices. No relationship is found between UV peak magnitude and the decline rate in contrast to what is typically found in optical band. The correlations found here are promising, and give encouraging insights for the use of SLSNe as cosmological probes at high redshifts using standardising relations in the UV. We also highlight the importance of early, and consistent, photometric data for constraining the light curve properties.
\end{abstract}

\begin{keywords}
supernovae: general -- cosmology: distance scale -- ultraviolet: general
\end{keywords}



\section{Introduction}
\label{sec:intro}
The newer generation of wide-format time-domain surveys over the past 15 years have discovered a rare class of highly luminous transients termed ``superluminous supernovae'' (SLSNe). These events are 10--100 times brighter at peak compared to classical type Ia and core-collapse supernova events, with total radiated energies of about $10^{51}$ ergs  \citep[e.g.,][]{smith2007,pastorello2010,galyam2012,quimby2018,angus2019}. SLSNe are exceptionally blue events and are characterised by slowly evolving light curves that remain optically detectable, within several magnitudes from peak, for 100s of days. SLSNe (type I) have been observed to have a preference to occur in low-metallicity, star-forming dwarf galaxies \citep{lunnan2014,leloudas2015,perley2016,schulze2018,hatsukade2018}. A recent comprehensive review of SLSNe is given by \cite{galyam2019}.

Previously, SLSNe were primarily defined by an arbitrary peak absolute magnitude cutoff of $M < -21$ mag in optical filters \citep{galyam2012}, however, fainter events have since been discovered which show similar spectroscopic and photometric behaviour \citep{dacia2018,lunnan2018,angus2019}. Therefore, this limit has been relaxed and SLSNe are now identified based on their unique spectral properties \citep{quimby2018}.  SLSNe are UV-luminous explosions, with majority of their spectral energy distribution (SED) emitted in the UV. This, combined with their extreme luminosities, makes SLSNe detectable upto very high redshifts (to $z\sim20$) with current and upcoming optical and near-infrared space- and ground-based telescopes.  Moreover, studies detecting SLSN at $z\sim$ 1.5--4 suggest that their rate is higher than that at lower redshifts \citep{niell2011,Cooke2012,howell2013,prajs2017}. Therefore SLSNe offer an appealing tool to study the high redshift Universe. 

Investigations of SLSN rest-frame UV light curves and key features found in their UV spectra help to understand their progenitors, explosion mechanisms, and physics behind their enormous energies.  Moreover, high redshift SLSNe and their rest-frame UV emission enable to investigate  both stellar and galactic evolution and physics. SLSNe are typically brighter near peak magnitude than their host galaxies and are one of our only means to probe the $z\gtrsim10$ Universe. As bright background beacons, SLSNe provide internal probes of their host proto-galaxies and the intervening material in absorption in the circumgalactic medium (CGM) and the intergalactic medium (IGM) along the line of sight. SLSN detections can trace star formation in high redshift dwarf galaxies (and potentially arrest star formation), which are believed to have contributed the most to cosmic reionisation. SLSN number-counts in well-defined volumes, as is done with SLSN detection, can place direct constraints on the high-mass end of the stellar initial mass function. SLSN ejecta and pre-explosion mass loss lend insight into the chemical enrichment of their host galaxies and the CGM and IGM over cosmic time. Finally, the very high redshifts at which SLSNe can be detected enable the study of the deaths of Population III stars, while providing our best chance to detect pair-instability supernovae. Here, we explore another potential utility of SLSNe as standardisable candles to probe the Universe from $z\sim$ 0--20. 

Reviewing the diversity in the population to date, SLSNe have been broadly classified into two classes \citep{galyam2012} based on their optical spectroscopic and photometric properties. SLSNe type I (SLSNe-I) are hydrogen poor and exhibit a blue continuum, with a distinctive ``W"-shaped or `comb'-shaped feature from OII absorption around $\sim 4200$ \AA\ during early epochs. At later times, SLSNe-I spectrum transforms to a SN-Ic like spectrum \citep{pastorello2010,quimby2011}. SLSNe type II (SLSNe-II) on the other hand show hydrogen emission lines and are likely related to Type IIn supernovae. The energy source for most SLSNe-II has been modelled as ejecta interaction with hydrogen-rich circumstellar material \citep[CSM;][]{smith2007,ofek2014,benetti2014,inserra2018a}. However, the power engine of SLSNe-I remains under debate, as radioactive decay of several solar masses of nickel fails to fully explain their light curve evolution and points to additional central energy input. Some proposed mechanisms include; central engine models, such as a magnetar spin-down \citep[e.g;][]{kasenbildsten2010,woosley2010}, the pair-instability process for stars with massive cores \citep{kasen2011,kozyreva2017}, and fallback accretion into a black hole that can also explain observed light curve undulations \citep{dexterkasen2013,kasen2016}. Another interesting feature that has recently been brought into light by \cite{Nicoll2016} is the presence of a small `bump' before the main peak in the light curves of some of the observed SLSNe-I. \citep[e.g.;][]{leloudas2012,nicholl2015,smith2016,angus2019}. In this work, we focus on SLSNe type I including those events with a pre-peak bump.

Although the physical understanding of the SLSNe-I explosion mechanisms and progenitor scenarios is still emerging, their fairly homogeneous observational behaviour has attracted significant attention for their potential use as cosmological probes for the local to high redshift universe \citep{king2014,inserra2014,wie2015,scovacricchi2016,inserra2020}. Driven by the fact that they show a relatively small dispersion in their peak optical magnitudes \citep{quimby2013}, SLSNe-I have recently been proposed as standardisable distance indicators to constrain cosmological parameters. \cite{inserra2014} [hereafter, \citetalias{inserra2014}] studied this prospect for the first time with a sample of 13 SLSNe-I over the redshift range of $0.1 < z < 1.2$ to develop a method of standardisation analogous to SNe Ia. They find a linear relationship between the peak absolute magnitude and decline rate of the light curves (over 10, 20, and 30 days after peak), measured in a synthetic filter centred at 400 \si{\nm}. This correlation reduce the scatter in peak magnitudes from $\sim 0.4$ mag to around 0.25 mag \citep[see also][]{papadopoulos2015}. They also find a similar relation with the change in colour of SLSNe over 30 days after maximum. This work gives a promising proof-of-concept that SLSNe-I could be standardised for measuring distances and larger data samples could increase the accuracy to be competitive with SNe Ia. However, \cite{dacia2018} explored similar correlations with various light curve properties of their sample of SLSNe-I but they did not confirm the above results. Recently, \cite{inserra2020} [hereafter, \citetalias{inserra2020}] built an updated sample and used a novel technique that classifies SLSNe-I based on their photometric properties in a 4-dimensional parameter space \citep{inserra2018}. With this more homogenised sample, \citetalias{inserra2020} obtained similar scatters as \citetalias{inserra2014} for the decline rate-magnitude and colour-magnitude relationships, thus further encouraging the exploration of SLSNe as standardisable candles. 

Overall, attempts to standardise SLSNe have been promising, however, current small data samples greatly hamper these efforts. Larger data sets are necessary not only to reduce statistical errors but also to understand the population diversity and their physics to consequently reduce systematic uncertainties. \cite{scovacricchi2016} showed that even an addition of $\sim$100 SLSNe-I to current SNe Ia samples could significantly improve the cosmological constraints by extending the Hubble diagram into the deceleration epoch of the Universe (i.e. $z > 1$). Upcoming transient surveys are expected to significantly increase the numbers and the redshift range of the detected SLSNe. \citetalias{inserra2020} predict detection of $\sim$900 SLSNe-I with the Vera C. Rubin Observatory\footnote{Legacy Survey of Space and Time (LSST) of the Vera Rubin Observatory is a $8.2$ m ($\sim$6.7 m effective) diameter telescope with $9.6 \mathrm{deg}^2$ field-of-view and will conduct several 10-year wide-field surveys across the Southern hemisphere in the \emph{ugrizy} filters.} in optical-NIR filters, (\emph{ugrizy}) to redshift $z \sim 4$, with a majority of the detections around the redshift 2. \cite{villar2018} also did simulations of SLSNe-I for LSST and predict a 10 times larger number ($10^4$ SLSNe per year) with most (90\%) detections at $z \lesssim 3$. It is expected that such large samples would constrain $\Omega_M$ and $w$ to 2\% and and 4\% respectively \citep{scovacricchi2016}. \cite{inserra2018euclid} show that the \emph{Euclid} satellite\footnote{\emph{Euclid} is a 1.2 m optical and near-infrared (NIR) satellite (550 - 2000 \si{\nm}) designed to probe the early Universe \citep{euclid2011}. Euclid Deep Survey (EDS) is particularly suited for long SLSNe light curves.} should detect approximately 140 (lower limit) high-quality SLSNe-I to $z \sim 3.5$ over the first five years of the mission.  The Nancy Grace Roman Space Telescope\footnote{Roman Space Telescope is an infrared observatory based on a 2.4 m primary mirror. One of its two instruments is the Wide-Field Instrument (WFI) that has a 300-megapixel infrared camera giving it a field of view which is a hundred times larger than the Hubble Space Telescope} will also perform deep wide IR surveys that can detect SLSNe to $z\sim$ 13.  Finally, the high sensitivity and long wavelength range of the \emph{James Webb Space Telescope} (JWST) could conduct a powerful survey for high-redshift transients and enable SLSNe detections to $z\sim 20$ in the near- and mid-IR bands \citep[e.g.,][]{wang2017}.  However, since JWST has a relatively small FOV, it will be highly useful for acquisition of spectra of high redshift SLSNe (and their host galaxies) detected by other facilities.

For the redshifts investigated here, i.e., $z\sim$ 1--3, that span the peak of cosmic star formation, current and future optical and NIR surveys will observe the SLSN rest-frame far-UV (FUV) and near-UV (NUV) emission. For example, the optical filters \emph{g, r} and \emph{i} of the Rubin Observatory will detect $\sim$ 160 \si{\nm}, 200 \si{\nm} and 250 \si{\nm} central rest-frame wavelengths, respectively, for an object at $z=2$. At higher redshifts, even NIR telescopes will sample rest-frame UV wavelengths for $z \gtrsim 6$ SLSNe. In terms of using SLSNe as cosmological probes, the major advantage is their detectability at redshifts beyond those possible for SNe Ia, i.e., redshifts $\gtrsim 1.5$. Thus, not only would SLSNe be used as a secondary check on SNe Ia results at $z\lesssim2$, but SLSNe would complement SNe Ia by extending the Hubble diagram to potentially $z\sim10$ and higher, enabling a distinction between various dark energy models well past the deceleration epoch. Therefore, perhaps the most powerful use of SLSNe for cosmology, using data from future instruments, lies at higher redshifts where they will be observed at their rest-frame FUV/NUV wavelengths. In summary, in order to exploit SLSNe for studying stellar and galaxy formation and evolution, and for their potential use as cosmological probes, it is crucial to characterise their UV behaviour. 

In this work, we investigate correlations among SLSNe-I UV light curve properties and explore their use as standardisable candles for the high redshift Universe. Past works \citepalias{inserra2014,inserra2020} investigated SLSN peak magnitude correlations with their decline rates and colours with a \SI{400}{\nm} filter. This work primarily focuses on the rising properties of SLSNe light curves, such as the rise time and colour evolution during rise, and explores corresponding luminosity correlations in the rest frame UV synthetic filters. 

There are a few physical and practical motivations to explore the rising part of the light curve instead of (or in addition to) the decline for standardising relations. Firstly, with the assumption that there is some uniformity in the underlying physics of SLSNe-I, one might expect more consistent evolution of the light curve at epochs right after the explosion and expansion compared to post-peak magnitude, where potential interaction with circumstellar material and unknown mechanisms (e.g., the SLSNe exhibiting post-peak bumps and undulations), effects/efficiency of magnetar energy transfer, and other aspects could make the light curve more complicated. Secondly, since we attempt to study the evolution properties of SLSN at bluer wavelengths, the effects from dust creation might be smaller at early times. Thirdly, work on SNe Ia standardisation have examined the peak magnitude correlations with the rising part of the light curve and have indicated its importance for cosmological measurements \citep[e.g][]{hayden2010,firth2015,zheng2018,hayden2019}, making it worthwhile to explore similar relations for SLSNe-I standardisation.  Fourthly, we aim to use a phenomenological approach because of the lack of understanding behind the explosion mechanisms and details behind SLSNe.  Finally, the limited non-uniform sampling and fragmented nature of the light curve data does not permit a study of the decline beyond $\sim$15 d for most cases for SLSNe at high redshift. Nevertheless, we do explore the decline part with the limited data, thus enabling a search for relationships between rise and decline with respect to peak magnitudes.

Besides the reasons presented earlier for exploring SLSNe in the UV for their cosmological use, characterising their blue wavelength behaviour is also important more generally for their detection, to help understand their explosion physics, their nature, and classification. For example, \citetalias{inserra2020} measured the pseudo equivalent width (pEW) of the C {\footnotesize III}/ C {\footnotesize II}/Ti {\footnotesize III} and Mg {\footnotesize II}/C {\footnotesize II} blended lines at $\sim$ \SI{2200}{\angstrom} and $\sim$ \SI{2800}{\angstrom} respectively, to search for a more quantitative way of distinguishing between Fast and Slow SLSNe detected at high redshifts. 

In this work, we assemble all the available UV and NUV SLSNe-I data from the literature, and include all events with data coverage on the rising part for our analysis. We measure their light curve properties and determine peak magnitude correlations in order to probe their potential for cosmological use. Given the presence of SLSNe with pre-peak bumps in their light curves \citep{Nicoll2016} and the debate over their ubiquity, we identify them separately within our data sample to look for differences, if any, in their trends from those of a general SLSNe-I sample. The intention is to help determine if there is different physics driving these explosions and whether or not they are a distinct population, providing insight on their nature. Additionally, since these events can be identified via photometry alone, we explore the prospect of using them as standardisable candles, or a population that can be easily eliminated from the full SLSN population, to enable `clean' photometric events for standardisation. Although the number of objects involved in this study is statistically small and the available data is sparse, this work provides an important first investigation of the evolutionary properties and potential cosmological use of the UV light curves of SLSNe.

We describe the SLSN data sample used in this work in Section \ref{sec:data}. In Section \ref{sec:method}, we outline the methodology for light curve fitting and present the estimated light curve parameters along with explaining various analytical techniques used to determine the correlation relations. Section \ref{sec:results} \ presents the observed relationships for various light curve properties of the two data samples. Finally we discuss our findings and draw the conclusions in Section \ref{sec:conclusion}


\section{Data}
\label{sec:data}
The primary goal of this work is to investigate the rest-frame UV behaviour of SLSNe-I light curves. The subset of existing high redshift SLSN data which includes rest-frame UV wavelengths is relatively small. From among the published events, we select all the SLSNe-I having rest-frame FUV/NUV photometric coverage with sufficient data to reliably measure their light curve properties (such as peak magnitude, rise time, colour evolution, etc.). This data set is referred to as the {\it `Literature'} sample because it includes all objects published to date that have photometric data in the rest-frame UV. Among this sample, we select SLSNe which pass certain data completeness and quality cuts in order to define a data set which allows us to measure the light curve properties with no, or negligible, extrapolations. This sub sample is called the {\it `SLSN-UV test'} sample and all correlations presented in this work are measured using only this sub-sample. The literature sample objects are shown only for completeness and comparison purposes. Additionally, as mentioned earlier, a lesser understood phenomenon observed in some SLSNe is the presence of a pre-peak bump in their light curves. We separately identify and mark such objects in our data sample, and refer to them as {\it `Bumpy'} SLSNe. Below, we describe in detail our data set.

\subsection{The Literature Sample}
\label{sec:slsnUV_data}

We take all published SLSNe type I from the literature having $z \gtrsim 1$ redshifts in order to assemble all SLSNe with available FUV/NUV data. This redshift cut is chosen such that the observed optical filters are blue-shifted to UV filters in the rest frame of the SLSN ($\lambda_{eff} = \lambda_{obs}/(z+1)$; for example, a SLSN at the lowest redshift of $z = 1$, the \emph{g} band effectively samples the spectral region around \SI{2400}{\angstrom}). Exceptions here are the lower redshift events that have UV data coverage with space-based telescopes such as the \emph{Neil Gehrels Swift Observatory (Swift)}. Secondly, in order to explore the rising phase of the light curve, we require photometric coverage from several days before the maximum to post maximum so as to efficiently constrain the rise time and the peak magnitude/epoch. Thirdly, often the data is not available in all optical filters, so we require data specifically in those observer-frame filters that coincide with UV/NUV bands in the SLSN rest frame. 

Finding a statistically significant sample which passes all these three criteria is challenging since there are only a few tens ($\sim 30$) of detected SLSNe-I at high redshift, and even fewer with adequate photometric coverage during the rising phase. Although acquiring high-cadence light curves for events at higher redshifts may be easier than low redshift owing to their slow evolving nature combined with the effect of time dilation, surprisingly few objects have coverage to later times (+30 days and beyond). Detecting these events on their rise is demanding because of their faintness at such high redshifts, and the fact that the temporal coverage of surveys with sufficiently deep and wide fields is comparatively sparse. High resource outlay also often prevents getting early spectral confirmation of these high redshift events since exposures of several hours on highly competitive 8m-class telescopes are generally required for the m$_r \gtrsim$ 24--25 near-peak magnitudes of SLSNe at $z \gtrsim 2$ \citep{smith2018,curtin2019}. 

Another challenge associated with SLSNe cosmology, as pointed out by \citetalias{inserra2020}, is to find a uniform classification scheme which relies on SLSNe progenitor scenarios and explosion mechanisms. Building a homogeneous sample is important in context of using SLSNe as standardisable candles because variations in the underlying physics of SLSNe may increase the intrinsic scatter in the correlations. In the absence of a robust definition of SLSNe-I sub-classes and understanding of their physics and explosion mechanisms, we do not make any distinctions in our data sample based on the light curve evolution of the SLSNe, unlike \citetalias{inserra2020}, where the objects are categorised into Fast and Slow types. We build the sample used in this work based solely on the availability of the photometric data and the required redshift cut ($z\gtrsim 1$ to reach the rest-frame UV/NUV), and study their light curve properties as a whole. The 
SLSN-UV test sample (see below) is also selected based only on the data quality and cadence. Furthermore, objects with early peaks \citep{nicholl2015} are not excluded if they pass our redshift and data quality criteria, instead, we identify them as a separate sub-set (see Section~\ref{sec:bumpy}).

Among the $\sim 30$ published high redshift events, there are 22 SLSNe which pass our three filtering criteria: (1) spectroscopic redshifts of $z \gtrsim 1$, (2) data coverage on the rising phase of the light curve, and (3) photometric coverage in the rest-frame UV/NUV filters, and hence these form our Literature Sample. An exception to the imposed redshift cut are the two events SN2017egm and SN2015bn at $z = 0.030$ and $z = 0.114$, respectively. These two objects have UV data coverage with the \emph{Swift} Ultraviolet and Optical Telescope (UVOT). Figure~\ref{fig:redshift} shows the redshift distribution for the Literature Sample. The correlations presented in this work are built using only the SLSN-UV test sample which is a sub-set of the Literature Sample (Section~\ref{sec:golddata}).

The Literature Sample is composed of events discovered and followed-up from several surveys. 
Of the 22 SLSNe-I in the Literature Sample, 8 are discovered with the Dark Energy Survey \citep[DES,][]{des2016}, an optical imaging survey using the Dark Energy Camera \citep[DECam,][]{flaugher2015} on the \SI{4}{\metre} Blanco Telescope at the Cerro Tololo Inter-American Observatory (CTIO) in Chile. These SLSNe were discovered during the DES-SN programme \citep{kessler2015,diehl2018} which surveyed 10 DECam pointings, imaging 27 deg$^2$ in $g, r, i$ and $z$ filters with an approximate 7-day cadence.These objects are reported and analysed in \cite{angus2019}.

Another 6 objects come from the Pan-STARRS1 Medium Deep Survey \citep[PS1 MDS,][]{chambers2016}, using the PS1 telescope on the summit of Haleakala in Hawaii, a wide-field survey instrument with a \SI{1.8}{\metre} primary mirror. PS1 MDS observed in $g_{P1}, r_{P1}, i_{P1}, z_{P1}$ filters with a typical 3-day cadence. These PS1 SLSNe are presented in various papers including \cite{chomiuk2011, mccrum2015} and \cite{lunnan2018}.

Three of the very distant SLSNe ($z \gtrsim 2$) in our sample are discovered with the Subaru HIgh-Z sUpernova CAmpaign (SHIZUCA) that uses the Hyper-SuprimeCam \citep[HSC,][]{Miyazaki2018,Kawanomoto2018} on the \SI{8.2}{\metre} Subaru telescope on the summit of Maunakea in Hawaii. HSC has a 1.8 deg$^2$ field of view and the necessary sensitivity required to find high-redshift SNe. Type-I classification of the three HSC SLSNe is uncertain owing to their low signal-to-noise spectra and have been assumed to be SLSNe-I for this work. These objects have been reported in \cite{moriya2019}.

Two SLSNe-I are included from the Supernova Legacy Survey \citep[SNLS,][]{perrett2010} that was based on the Deep Survey of the Canada-France-Hawaii Telescope Legacy Survey \footnote{\href{https://www.cfht.hawaii.edu/Science/CFHLS/}{Canada-France-Hawaii Telescope Legacy Survey}} (CFHT-LS). CFHT-LS Deep Fields imaged four fields in $g, r, i$ and $z$ filters with a cadence of 3-5 days. These are presented and analysed in \cite{howell2013}. Additionally, another CFHT object in the sample is SLSNe SN2213-1745 ($z = 2.046$), discovered by \cite{Cooke2012} in the CFHT-LS Deep Fields using an image stacking technique. We note that this work discovered another SLSN, SN1000+0213, at $z \approx 4$ which is the highest redshift SLSN detected to date and also exhibits a pre-peak bump. However, this object is excluded from the SLSN-UV sample because the ground-based optical date probes too blue (rest-frame {\it gri} data coverage of \SI{850}{\angstrom} to \SI{1700}{\angstrom}; z-band data are too shallow for this work) compared to the rest of the sample.

Finally, two low redshift SLSN events; SN2017egm at $z = 0.03$ and SN2015bn at $z = 0.114$. SN2017egm was discovered by the \emph{Gaia} Satellite (presented in \cite{nicholl2017}) and SN2015bn (presented in \cite{nicholl2015b}) was first discovered by the Catalina Sky Survey. These SLSNe are included in our sample as they have rising phase data coverage with the \emph{Swift}-UVOT UV filters. 

Table \ref{tab:table1} lists all the 22 objects that comprise the Literature Sample along with their redshifts and photometric data references. Redshifts of all the SLSNe-I used in this work were determined spectroscopically, with the exception of HSC16apuo, where the host galaxy redshift is estimated photometrically as a distribution between $z \simeq$ 2.8 and 3.5 \citep{moriya2019}, with the most probable value from that work, $3.22$, adopted here.

\begin{table*}
\centering
\caption{Literature Sample consisting of 22 SLSNe-I. \textit{Column 3} lists the observed filters used for calculating the magnitudes in the \SI{250}{\nm} and \SI{310}{\nm} synthetic filters. \textit{Column 4} and \textit{Column 5}  give the rest frame coverage of the filters used for \SI{250}{\nm} band and \SI{310}{\nm} band respectively. \textit{Column 6} specifies whether or not the SLSN is part of the SLSN-UV test sample (i.e. the sub-sample used for measuring the correlations), and \textit{Column 7} tells if the SLSN shows a pre-peak bump in their light curves. Literature reference -- a) \citet{angus2019}, b) \citet{chomiuk2011}, c) \citet{lunnan2018}, d) \citet{mccrum2015}, e) \citet{moriya2019}, f) \citet{howell2013}, g) \citet{Cooke2012}, h) \citet{nicholl2017}, i) \citet{Nicoll2016}}
\input{tables/Table1_redo}
\label{tab:table1}
\end{table*}

\subsubsection{The SLSN-UV test sample: SLSNe used here for testing correlations}
\label{sec:golddata}
Given the poor physical understanding of SLSNe and their classifications, coupled with the paucity of UV data, selection criteria adopted here to build the literature data sample rely more on the availability of the UV data than on the physical properties of the SLSNe. However, not all the SLSNe-I available in the literature have data quality which allows us to measure light curve properties with small uncertainties and without biases on the light curve shape. For example, no early data to measure the rise time or missing peak data because of limited observing seasons visibility during the year. This happens particularly for high redshift SNe where time dilation stretches the transient's visibility in the observer frame such that the full light curve evolution (rise an decline) is not covered in one observing season. 

Taking into account the data quality and cadence, we select SLSNe from the Literature sample that have (1) a well defined main peak with at least one data point before and after maximum and (2) rising light curve data from at least 1 magnitude before maximum so that the rise time can be measured reliably. These criteria are used to enable peak magnitude and rise time measurements with negligible extrapolations. Thirteen of 22 SLSNe from the Literature sample pass the above criteria and comprise the SLSN-UV test sample (Table \ref{tab:table1}), which is used for all the cosmological correlations explored in this work.

\subsubsection{Bumpy SLSNe}
\label{sec:bumpy}
Several SLSNe detected to date have been observed to exhibit multiple peaks in their light curves. Some objects show re-brightening at later times during the decline of the main peak, and others (with sufficient early and deep data) have been observed to exhibit bumps prior to the main peak of the light curve \citep[e.g.,][]{leloudas2012,nicoll2015,smith2016,anderson2018}. While some SLSNe show confirmed pre-peak excess in their photometry, it is unclear whether this feature is common among the full SLSNe-I population because pre-peak bumps are faint and may fall below the detection limits of the surveys or individual observations. While \cite{Nicoll2016} argue that bumps could be ubiquitous, \cite{angus2019} concluded contrary with DES SLSNe data analysis. The physical mechanism behind the early bumps also remains unclear due to the lack of spectroscopic information during the bump epoch. Studying the nature of \emph{bumpy}\footnote{\emph{Bumpy} in this works refers to bumps before the main peak of the light curve and not to the late-time undulations} SLSNe may provide the key toward understanding the initial explosion conditions and progenitor scenarios of SLSNe and whether these events are a separate population.

The physical understanding of pre-peak bumps notwithstanding, the distinct photometric signature of the these SLSNe would make them powerful tools for cosmological use. As such, a distinguishing feature enables SLSN identification without spectroscopy, in particular at high redshift, where spectroscopy is very time-expensive. Homogeneity in the underlying physics could help explain their light curves and help their standardisation. Therefore, owing to their characteristic light curve behaviour, we separately identify events in our Literature sample which have a bump (or a possible bump) before the main peak in their light curve and refer to them as \emph{`bumpy'} SLSNe (or \emph{`may-be bumpy'} SLSNe for objects with hints of a possible early bump). We note that `bumpy' here refers to the objects which have a confirmed pre-peak bump in their light curve data, while `may-be-bumpy' refers to the objects where the data suggests the presence of pre-peak excess however not deep and early enough to confirm it. This separate sub-identification may help discern any trends associated with these bumpy SLSNe in the various correlations explored here. Objects that do have neither any signature nor a possibility of a pre-peak bump in their data are not included in either of the bumpy or may-be-bumpy categories and are simply assumed to have no pre-peak bumps. They are called 'non-bumpy' to differentiate them from the former.

There are 10 SLSNe among the full Literature Sample of 22 which either have a confirmed pre-peak bump (i.e. 4 bumpy SLSNe), or the possibility of a pre-peak bump in their light curves could not be excluded (i.e. 6 may-be bumpy SLSNe) and are listed in Table \ref{tab:table1}.

\begin{figure}
\centering
\hspace*{-0.5cm}
\includegraphics[width=7.0cm,height=7cm]{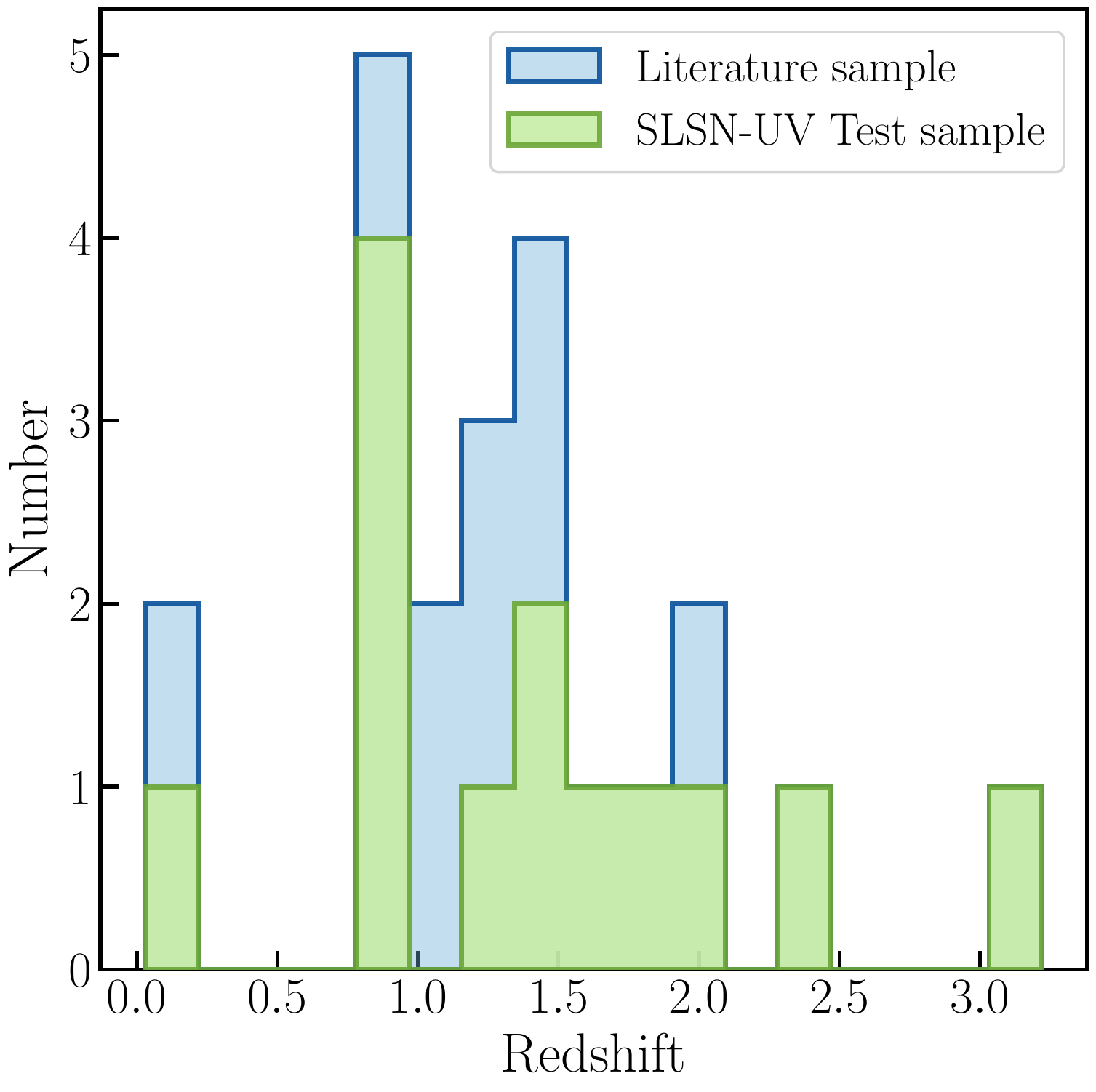}
\caption{Redshift distributions of the SLSNe in the Literature Sample (22 SLSNe) and the sub-set SLSN-UV Test sample (13 SLSNe).}
\label{fig:redshift}
\end{figure}


\section{Method}
\label{sec:method}
This work aims at testing standardisation of SLSNe in the UV to enable their use as cosmological probes from low to high redshift, in particular for $z \ge 1$. As such, for SLSNe at redshifts $z \approx$ 1--6, standard optical filters probe the rest-frame from $\sim$ \SI{2500}{\angstrom} to the far-UV. An immediate challenge is the small number of high redshift \slsn detected to date with well-sampled light curves in a specific optical filter. With this in mind, we attempt to develop an optimised framework that interpolates the observed SLSNe light curves at any epoch, even if they are sparse, and measures their properties and evolution. We aim to estimate the peak magnitude, the epoch of peak magnitude, and to characterise their colour and light curve behaviour. Furthermore, in order to compare peak magnitudes in a synthetic filter, we also explore a viable solution for \emph{K}-correcting the estimated peak magnitudes in the observed filters into a synthetic band in UV (described in section \ref{subsec:Kcorr}), in the absence of UV spectra at peak for each SLSN.

Apparent magnitudes at all epochs are converted to absolute magnitudes using flat $\Lambda$ cold dark matter ($\Lambda$CDM) cosmology with $H_0 = 70\  \mathrm{km\ s^{-1}\ Mpc^{-1}}, \Omega_{\Lambda} = 0.7$ and $ \Omega_{\mathrm{M}} = 0.3$. Correction for cosmological expansion is applied to all the absolute magnitudes, hence absolute magnitude $M$ at any observed epoch is given by $M = m - 5 \log_{10} (d_L/10pc) + 2.5 \log_{10}(1 + z)$. Light curves are corrected for time dilation and all timescales are given in the SLSN restframe throughout the paper. Photometric data has been corrected for Milky Way extinction \citep{schlafly2011}, however, no correction is applied for the extinction in SLSN host galaxies, which is assumed to be small \citep{nicholl2015b,leloudas2015,inserra2020}. Several factors support this assumption, including that SLSNe-I have been observed to typically occur in dwarf, low metallicity hosts \citep{lunnan2014,chen2017,perley2016,angus2016,izzo2018,hatsukade2018} and often on the outskirts of the galaxies in space-based imaging \citep[][see also \citealt{angus2016} and \citealt{lunnan2015}]{curtin2019}. Furthermore, low scatter has been observed in the SLSNe UV peak magnitude distribution \citep{smith2018} and in the colour distribution in the optical and UV \citep{inserra2018,smith2018}.

The uncertainties on the absolute magnitudes are directly propagated from the uncertainties on the observed apparent magnitudes. We do not include any errors from the distance estimates with the redshift and the assumed cosmological model, as they are negligible.

\subsection{Synthetic photometric bands}\label{sec:syn_bands}
\begin{figure}
\centering
\hspace*{-0.5cm}
\includegraphics[width=8.5cm,height=6.5cm]{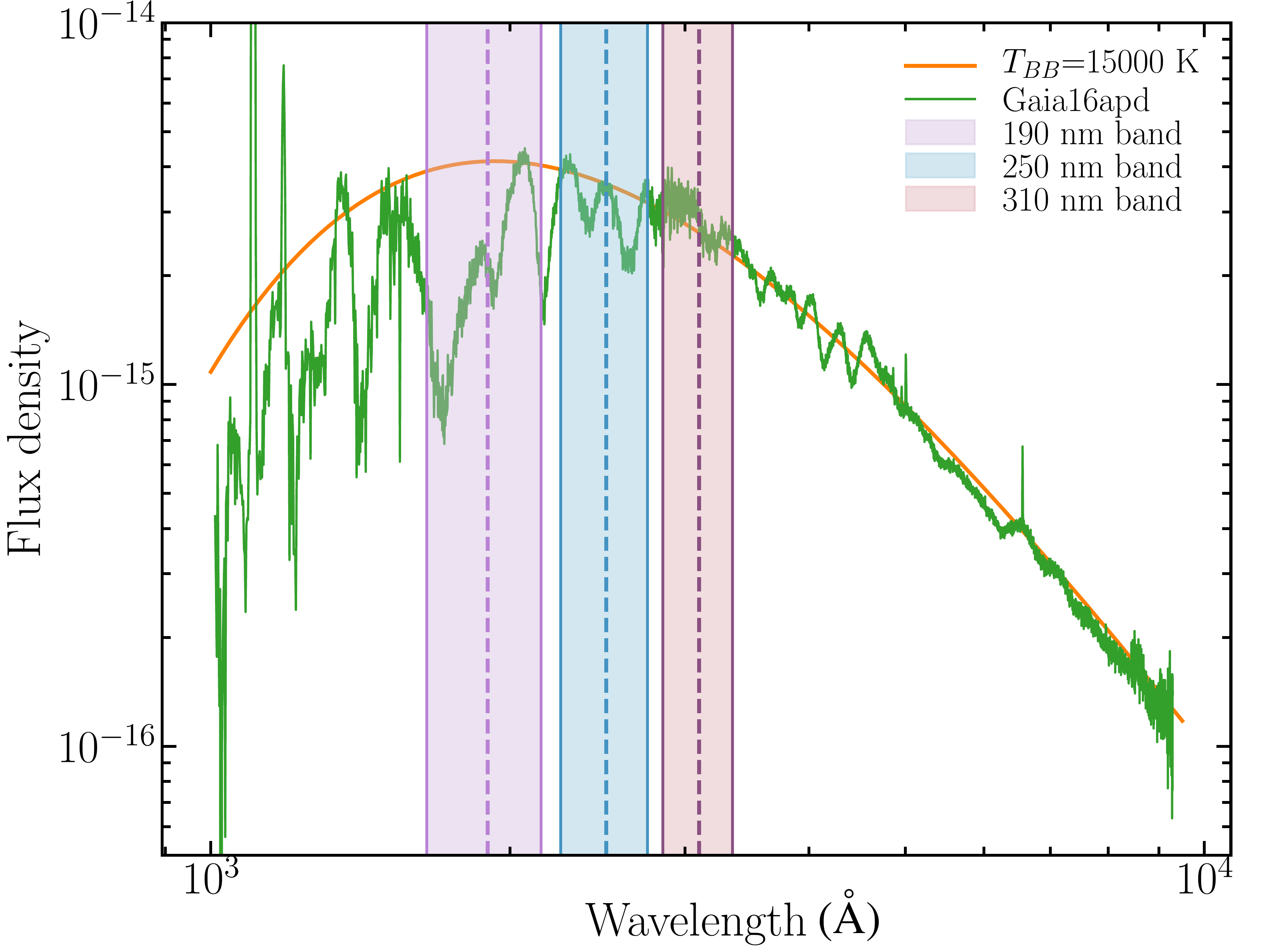}
\caption{Gaia16apd spectrum at peak \citep{linyan2017} and the blackbody function at 15000K. Also shown are the four synthetic filters, \SI{190}{\nm}, \SI{250}{\nm}, \SI{310}{\nm} and \SI{400}{\nm} used in this work.}  
\label{fig:spectrum}
\end{figure}

The intention here is to explore SLSN-I FUV/NUV peak magnitude correlations, however, the details of the spectra of  SLSNe-I in the UV are poorly known. Currently, \emph{Gaia16apd} is one of only two SLSNe-I that have UV spectral data at/near the peak light \citep{linyan2017,linyan2018} and we use it here as a template for the near-peak spectral behaviour of SLSNe-I to evaluate \textit{K}-correction (See section \ref{subsec:Kcorr}). Strong absorption features are observed in the rest-frame FUV and, although these absorption features are key to understanding SLSN physical processes, they could also introduce additional photometric scatter in SLSN light curves.  One would ideally want to explore light curve correlations in a continuum region devoid of strong absorption features, however, this becomes difficult blueward of $\sim$3000\AA\ (Figure \ref{fig:spectrum}). Below, we motivate our choice to use synthetic photometric bands, keeping in mind the above considerations and to maximise the sample size.

For our data sample, we build a synthetic passband with a width of \SI{500}{\angstrom} centred at \SI{2500}{\angstrom}. This band contains two moderately strong absorption features that appear relatively consistent in \slsn spectra obtained to date. This band is referred to as the \SI{250}{\nm} band throughout this work and is shown in Figure \ref{fig:spectrum}. We define another similar synthetic band centred at \SI{3100}{\angstrom} (Fig.\ref{fig:spectrum}), referred to as the \SI{310}{\nm} band, and it is used to compute colour (250-310) of the SLSNe. This synthetic band probes a relatively featureless portion of the spectrum and has good observed photometric coverage in the data, while still remaining in the NUV spectral region. Table \ref{tab:table1} presents the observed filters which are \emph{K}-corrected to corresponding synthetic filters for the whole Literature Sample.

As the aim here is to explore SLSNe in the rest-frame UV also for higher redshift objects for cosmology, we define a bluer synthetic filter with a width of \SI{500}{\angstrom} centred at \SI{1900}{\angstrom} (see Figure \ref{fig:spectrum}). This \SI{190}{\nm} band has large absorption features and only 6 of the highest redshift SLSNe have the required data.

\subsection{Light Curve fitting}
\label{sec:lcfitting}
In order to estimate the peak brightness of \slsn and interpolate the photometric observations to characterise their light curve behaviour, we employ Gaussian Process Regression (GPR) technique \citep{gpr1,gpr2}. Being a non-parametric Bayesian approach, this method removes any assumptions on the SLSN light curve shape, which may be introduced with polynomial fitting. Comparing polynomial fitting with Gaussian Process, \cite{inserra2018} show that a Gaussian Process fit represents the observed data better than a polynomial fit. GPR works well on small data sets common in transient astronomy and has the ability to provide uncertainty measurement intrinsically on the predicted values, which is particularly important in this work. 

This technique has the advantage over other methods by including uncertainty information of the observed data, thus producing less-biased interpolated values. Additionally, this method is very powerful for SN light curves having incomplete or noisy photometric data. Therefore, GPs are being successfully used in astronomy \citep{mahabal2008,way2009,gibson2012} and supernovae analyses \citep{mandel2009,kim2013,Scalzo2014,lochner2016}. The use of Gaussian Process techniques allows the user to marginalise over systematic sources of noise within a data, which might otherwise not be captured in an astrophysical model.

Gaussian Process (GP) is a probability distribution over all admissible functions that can model the correlated noise within a set of temporal or spatial data. It is a collection of possibly infinite random variables and hence termed non-parametric. Among these infinite parameters, any finite number set has a multivariate Gaussian distribution. Just like a Gaussian distribution, a GP is defined by (1) a mean function $\mu(x)$ which determines the mean at any point of the input space, and (2) a covariance function, a.k.a., `kernel' $K(x, x^\prime)$, which sets the covariance between data points $x$ and $x^\prime$.
\begin{equation}
    f(x) = GP(\mu(x), K(x, x^\prime))
\end{equation}

For any input point $x$, the reconstruction function $f(x)$ has a normal distribution with its mean value given by the mean function $\mu(x)$ and a covariance between two points by the kernel function $K(x, x^\prime)$. The kernel determines the kind of relationship between adjacent data points making one point dependent on the other. Both $\mu$ and $K$ may be parameterised and the parameters of the latter are referred to as hyper-parameters since they describe the function scatter rather than the function itself. There are two hyper-parameters; the vertical scale $\sigma$ that describes how much the function can span vertically, and the horizontal scale $l$ that tells how quickly the correlation between two points drops as the distance between them increases. A high $l$ gives a \textit{smooth} function, while a lower $l$ results in a \textit{wiggly} function. For time-series data as dealt with here, these two hyper-parameters respectively become the uncertainties in the measured magnitudes and the timescale over which significant changes occur within the data. The functional form of the covariance function or `kernel' used can be selected/constructed such that it represent any periodic tendencies within the data.

The likelihood function of a GP is a multivariate Gaussian distribution of dimension equal to the number of measurements $n$. The functional form of the covariance function defines the relationship between the measurements and it absorbs any intrinsic systematics which are unknown to the user. The kernel hyper-parameters may be optimised to reach the best convergence of the distribution.

We implement GPR to interpolate the observed light curves of all the SLSNe in our sample. We use a flexible python library called \texttt{GEORGE} \citep{Ambikasaran2015} and employ a Matern 3/2 kernel already implemented in this library. A Matern 3/2 kernel is mathematically similar to squared exponential function and can be written as:
\begin{equation}
    k(r) = \sigma^2 \left(1 + \frac{\sqrt{3}r}{l}\right) \exp \left(\frac{-\sqrt{3}r}{l}\right)
\end{equation}

where $\sigma$ is the vertical span \textit{i.e.,} error in an observation, $l$ is the horizontal scale over which variations happen in the data, and $r$ is the separation between observations.

This kernel provides a greater flexibility in fluctuations over short time scales and has been shown to best represent the form of SLSN light curves \citep{inserra2018, angus2019}. Our GPR fitting method determines the best fitting hyper-parameters for each light curve via a gradient based optimisation. We interpolate the light curves over the full temporal range of measurements and estimate their evolution parameters. The fitting is done for the chosen observed band(s) for each SLSN in the Literature Sample.

Figure~\ref{fig:UVlcfit} presents the mean and 1 $\sigma$ uncertainties for the GP fits of all 22 objects in the Literature Sample. The 13  SLSNe constituting the SLSN-UV test sample are marked with an orange star on their light curve plot. The figure shows photometric measurements for all observed bands for each SLSN, while the GP fit is shown only for one chosen filter which is closest to the \SI{250}{\nm} band (as given in Table \ref{tab:table1}) based on the SLSN redshift. For those SLSNe which have an early bump in their light curve, the data points in the bump phase (where present) are not included in the fitting unless the absence of sufficient data characterising an early bump does not skew the fitting process. For all GP interpolated fits, we observe that the light curves are less flexible or `tighter' in areas with high cadence data, while interpolations over large data gaps are more uncertain or `loose'. The optimised kernel defines the relationship between successive points and greater data density results in fewer degrees of freedom for the fit, while sparsely spaced data generate more uncertainty.

\begin{figure*}
\centering
\hspace*{-1.0cm}
\includegraphics[width=17.0cm,height=23.0cm]{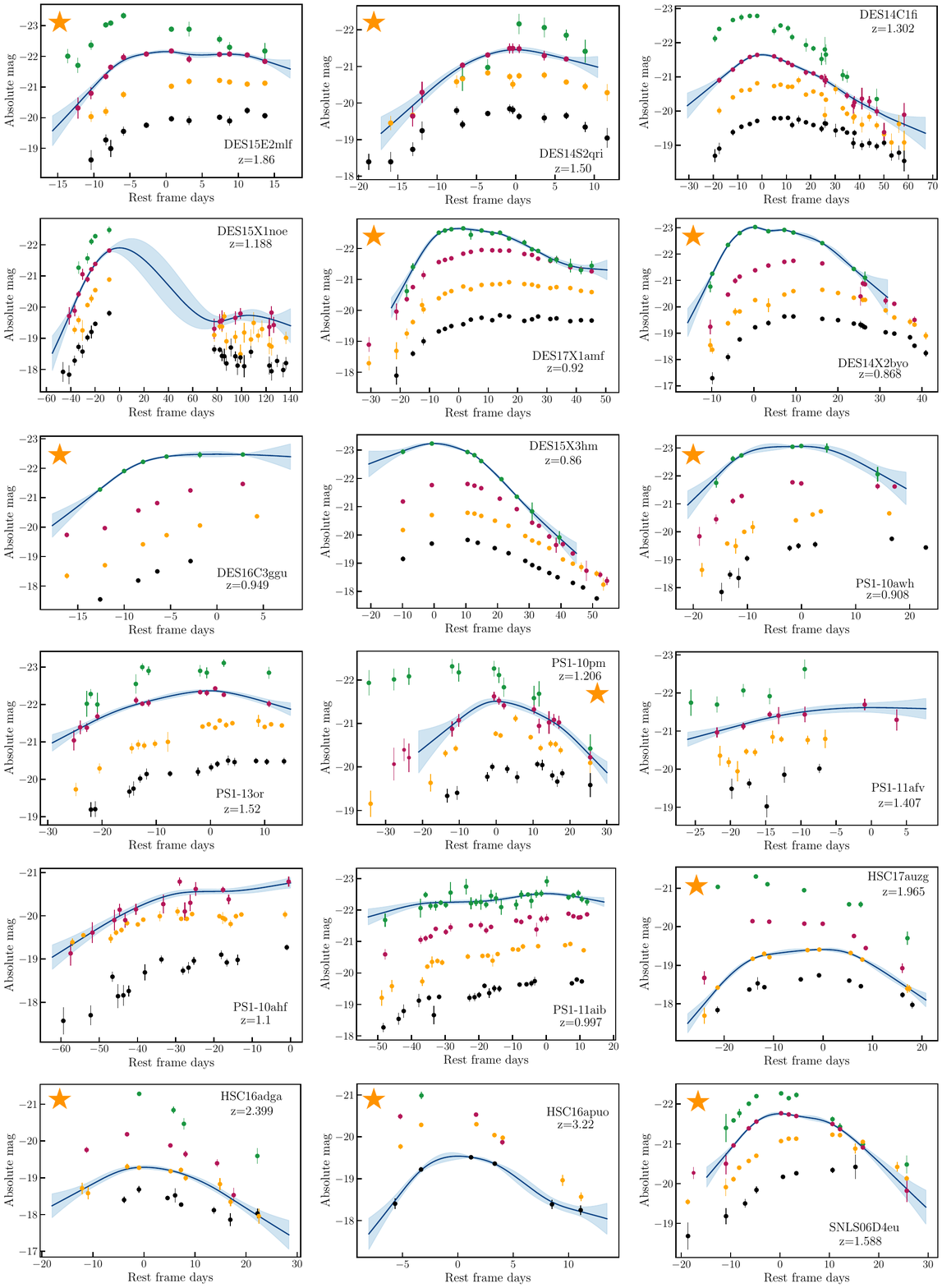}
\caption{Light curve fits for the 22 SLSNe in the Literature Sample. Objects marked with an orange star denote they are part of the SLSNe-UV test sample. All phases are given in SLSN rest-frame days relative to the peak magnitude in the fitted observed band. The bands are chosen such that they are closest to the \SI{250}{\mm} filter in rest-frame as given in Table \ref{tab:table1}. \emph{(continued on the next page)}}
\label{fig:UVlcfit}
\end{figure*}

\renewcommand{\thefigure}{\arabic{figure} Cont}
\addtocounter{figure}{0}

\begin{figure*}
\ContinuedFloat
\centering
\includegraphics[width=17.0cm,height=8.0cm]{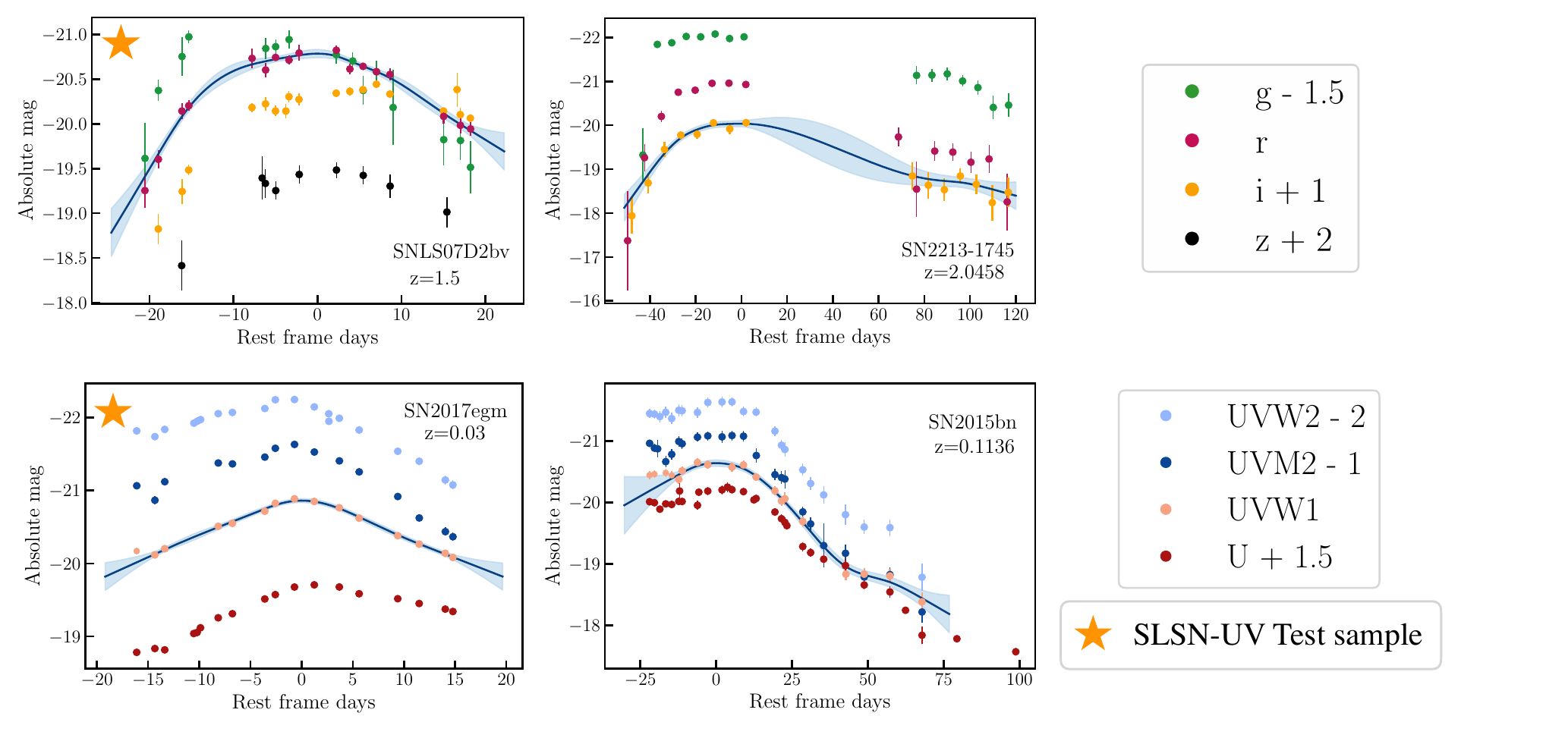}
\caption[]{Continued Figure \ref{fig:UVlcfit}.}
\end{figure*}
\renewcommand{\thefigure}{\arabic{figure}}

\subsubsection{Light Curve properties}
\label{sec:lcprop}
We can estimate SLSN magnitudes and uncertainties at any epoch from the GP fitted light curves, regardless of the data cadence. For each SLSN, we quantify its peak and evolutionary behaviour using certain quantities measured with the interpolated light curve. Below, we define each of these quantities along with their notation which are used throughout this work.

\begin{enumerate}
    \item \textit{Peak magnitude} - The maximum absolute magnitude in a photometric band $X$ written as $M_0(X)$, where $0$ indicates the peak epoch. For example, peak magnitude in \SI{250}{\nm} band is written as $M_0(250)$.
    \vspace{0.1cm}
    \item \textit{Rise time (1 mag to peak)} - The time in rest frame days that a SLSN takes to rise from 1 magnitude below peak to the peak magnitude. It is denoted as $\tau_{\mathrm{rise}}^{\Delta1 \mathrm{m}}$. Rise time here is measured in the \SI{250}{\mm} rest frame filter except in Figure \ref{fig:3100} where it is measured in \SI{310}{\mm} filter.
    \vspace{0.1cm}
    \item \textit{Rise rate} - The change in magnitude from 15 rest-frame days before the peak magnitude to the peak magnitude. The rise rate is denoted as $\Delta M_{-15}$.
    \vspace{0.1cm}   
    \item \textit{Decline rate} - The change in magnitude from peak magnitude to 15 rest-frame days after the peak magnitude. It is denoted as $\Delta M_{15}$
    \vspace{0.1cm}    
    \item \textit{Colour at peak} - The colour index of a SLSN at the epoch of peak magnitude and is defined as $M_0(X) - M_0(Y)$, written as $(X-Y)_0$. For example, the colour at peak magnitude in the \SI{250}{\nm} and \SI{310}{\nm} bands is denoted as $(250-310)_0$.
    \vspace{0.1cm}   
    \item \textit{Delta colour} - The change in colour from 15 days before peak magnitude to the peak magnitude, denoted as $\Delta (X-Y)_{-15}$. For example, $\Delta (250 -310)_{-15}$.
\end{enumerate}

All quantities are measured using the interpolated light curves. To measure the changes, the epoch of 15 days before peak is chosen because many SLSNe in our sample do not have very early data and estimating/extrapolating to earlier epochs would incur assumptions on the light curve shape. In few cases where there was no data 15 days before/after the peak in the required observer band (for example, HSC16apuo, SN2015bn), we use extrapolated values with associated uncertainties that are larger than those of interpolated values, and those objects are not included in the SLSN-UV test sample used to determine the peak magnitude correlations. Table \ref{tab:table2_slsnUV} lists the measured light curve properties for the full Literature Sample.

\begin{landscape}
\begin{table}
\centering
\caption{Light curve properties of the 22 SLSNe in the full Literature sample, measured with the interpolated GP fits. The peak magnitudes $M_0(250)$ given here have been \emph{K}-corrected using the \emph{Gaia16apd} spectrum at peak (see Section \ref{subsec:Kcorr}). The errors on all properties are estimated intrinsically with GPR, while errors on the rise time are computed using data resampling (see Section \ref{sec:timeerror}). The last column gives the SLSNe numbers used on some plots as aliases for their names.}
\label{tab:table2_slsnUV}
\input{tables/slsnUV_table2}
\end{table}
\end{landscape}

\subsubsection{Error on Rise time}
\label{sec:timeerror}
While the errors on interpolated magnitudes at peak or any other epoch are directly estimated by GP fitting depending on the cadence and uncertainties in observed data, the time stamp of any observed data has essentially zero error. Though the gaps in photometric data cause uncertainty in predicting the time of maximum, this uncertainty in peak epoch is not explicitly measured with the GPR. Quantifying the error on the estimated peak epoch is important since we want to measure the rise time (here, time taken to rise 1 magnitude below peak to the peak). We derive the uncertainties in peak epoch and the rise time by using a data resampling technique with Monte Carlo method \citep[See e.g.,][]{burtscher2009}.

\begin{figure}
\centering
\hspace*{-0.5cm}
\includegraphics[width=9.0cm,height=4cm]{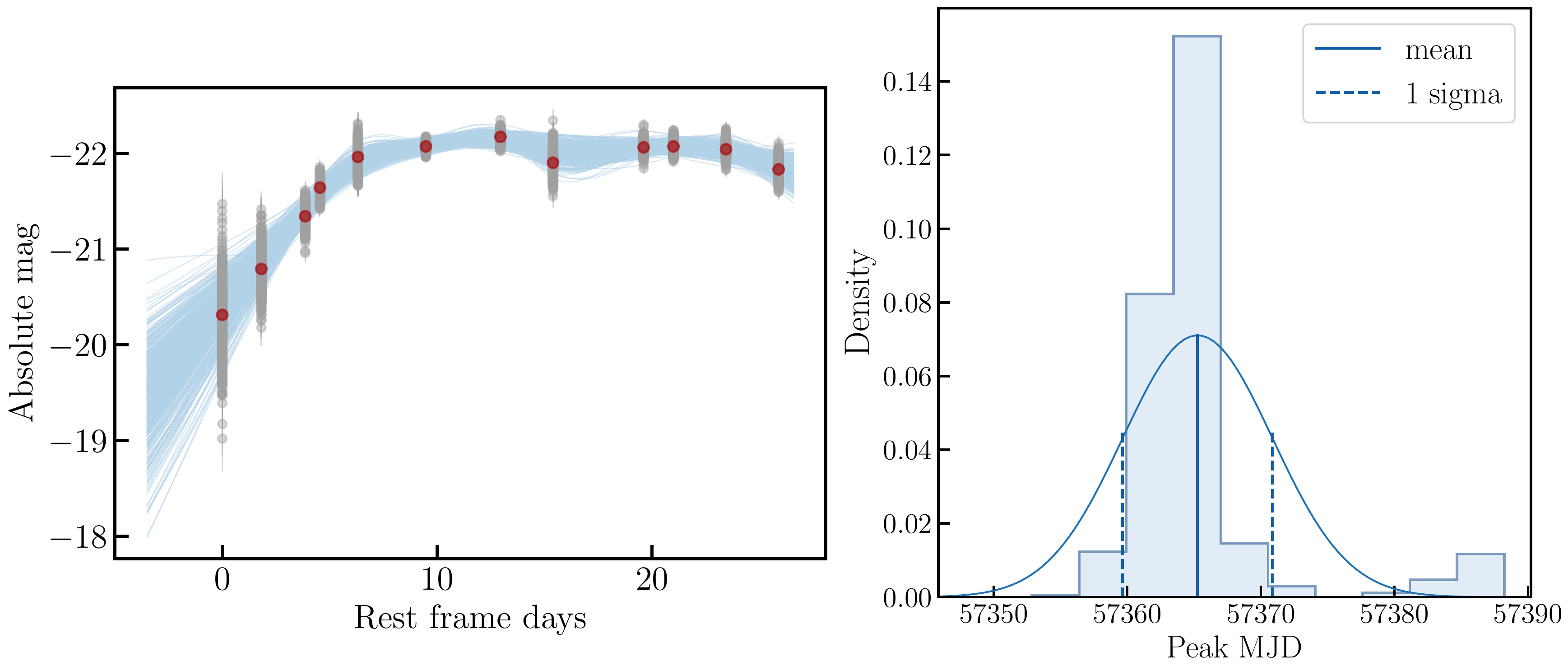}
\caption{Resampled light curves simulated from the measured data of SLSN DES15E2mlf are shown in blue (left panel) along with the density distribution for the measured peak epoch for each one these light curves (right panel). The error on peak epoch is estimated as the standard deviation of the distribution.}
\label{fig:errtime}
\end{figure}

This approach is intuitive to error estimation in the sense that it simulates repeated measurements of the light curve. We assume that the error distribution of the measured data is Gaussian. For each data point $x_n$ on the light curve, we invoke a Gaussian with mean $x_n$ and standard deviation $\sigma _n$ as the measured uncertainty. We randomly sample a new data point $x^\prime_n$ from this distribution. Similarly, for all data points, we generate an \textit{alternative} light curve with the same temporal sampling as the observed light curve. We then estimate the evolution parameters for this resampled light curve, namely peak magnitude, peak epoch and the epoch for 1 magnitude before peak using GPR. Repeating this resampling process 500 times, and estimating the light curve parameters each time, we get a frequency distribution of each parameter. We can then infer the uncertainty in that parameter as the spread of the frequency distribution. Figure~\ref{fig:errtime} presents an example of the resampled light curves (left panel) and the resulting distribution of the time of maximum (right panel) for SLSN DES15E2mlf. We estimate the errors on the peak epoch and on the epoch for 1 magnitude before the peak for each SLSNe in our data sample. These two errors are then added in quadrature to give the final error on the rise time. 

This method provides an upper limit of the uncertainty, because even though we use the measured error as the standard deviation of the Gaussian, we centre this error distribution on the measured value instead of the unknown true value. This introduces additional scatter and leads us to overestimate the uncertainty. Nevertheless, it still provides a conservative estimate of the error which is useful for the purpose of this work.  

\subsection{Cross-filter \textit{K} corrections}
\label{subsec:Kcorr}
A comparative study of peak magnitudes for a set of objects having a wide redshift range (See Figure~\ref{fig:redshift}) requires \textit{K}-correcting the peak magnitudes to a single common synthetic band. The method of \textit{K}-correction requires the spectrum of each object at or around the peak epoch. As described in section \ref{sec:syn_bands}, peak absolute magnitudes for our data sample are determined in the \SI{250}{\nm} and \SI{310}{\nm} (used for calculating colour) rest-frame bands. Additionally, in order to reach further blue in the spectrum, we compute the peak magnitudes also in the \SI{190}{\nm} band for higher redshift SLSNe.

In order to \textit{K}-correct the estimated peak magnitude of a SLSN in an observed filter into a fiducial UV filter, one requires the SLSN spectrum taken at or near peak epoch with coverage in the UV. We intend to use spectra at peak or within $\pm 5$ days from peak to calculate K-corrections. However, we found only 5 objects out of 13 in our SLSN-UV test sample ($<40\%$) which have a UV/NUV spectrum taken within $\pm 5$ days with none of them at peak. This number significantly limits any analysis. Therefore, we resort to \textit{K}-correcting our peak magnitudes by adopting either a `standard' template spectrum or a blackbody (BB) curve depending on the synthetic filter in question. SLSN spectra have been observed to be largely featureless redwards of $\lesssim 3000 \A$, and are well represented by a blackbody at those wavelengths. However, they show significant absorption features at UV wavelengths below $\lesssim 3000 \A$, that become even stronger shortward of $\sim2000 \A$. The strength of these UV absorption features is generally highest around SLSN peak magnitude (i.e., when the photosphere is hottest), and decreases as the supernova cools down \citep{angus2019}. 

As mentioned before in sec \ref{sec:syn_bands}, \emph{Gaia16apd} is one of two SLSNe-I that have UV spectra at/near the peak \citep{linyan2017,linyan2018}. We use this spectrum for \textit{K}-correcting our peak absolute magnitudes in the UV filters with the assumption that it represents a `standard' SLSN-I spectrum over this wavelength range \citep{smith2018}. Besides \emph{Gaia16apd} being the only UV spectrum at peak, using it as a template spectrum provides a baseline against which future measurements can be compared. Additionally, we minimize the \textit{K}-corrections as much as possible by choosing observer-frame filters closest to the synthetic-filters (as given in Table \ref{tab:table1}). We further test this assumption as given below to make sure that it does not compromise any results. Figure \ref{fig:spectrum} shows the \emph{Gaia16apd} spectrum at peak, the BB fit to the spectrum, and the three synthetic bands used in this work. 

Among our synthetic filters, the \SI{310}{\nm} band is the bluest wavelength range that has few absorption features and follows a BB function well. The \SI{250}{\nm} has two relatively moderate-strength broad absorption features. Shortward of $\sim$2200 \AA, there are a number of broad absorption features, including those near $\sim$2100, 1700, and 1400 \AA.  We note that although the absorption features strongly affect the flux as compared to a BB spectrum, measuring magnitudes over these features is useful if the features are shown to be consistent amongst SLSNe-I. Moreover, with a larger sample of near-peak spectra, the flux scatter in these UV bands can be computed and tested for their usefulness.  With the discovery of higher redshift SLSNe-I, one or more synthetic bands blueward of \SI{190}{\nm} and redward of Lyman-$\alpha$ may likely be required.

We adopt a hybrid approach for calculating \textit{K}-corrections to our synthetic filters. Since the \SI{310}{\nm} band (used for calculating color at peak) is continuum dominated and closely follows a blackbody, the K-corrections to the this band have been computed using a constant temperature BB spectrum of \SI{15000}{\kelvin}. This value was adopted as a mean value among the various SLSN-I peak temperatures evaluated in the literature. To verify this, we also measure the peak temperatures of all SLSNe in our sample using the available photometry and found them to be scattered around \SI{15000}{\kelvin}. Comparing K-corrections computed using individual temperature BB curves with those computed using 15000K BB curve, we find the differences to be smaller than the individual errors on the peak absolute magnitude. Additionally, owing to the high redshifts of the objects, majority of the photometry is in NUV rest frame bands (~\SI{200}{\nm} -- \SI{450}{\nm}) where spectral region bluer than \SI{300}{\nm} has strong absorption features. Hence, fitting BB to this blue photometry does not give accurate results. Arbitrarily removing bluer bands data from the BB fits often leave us with two or even one data which is not enough for fitting. Hence, for consistency and minimising arbitrary assumptions, we decide to use a constant temperature BB curve at \SI{15000}{\kelvin} for calculating the K-corrections to the \SI{310}{\nm} band.

\begin{figure}
\centering
\hspace*{-0.5cm}
\includegraphics[width=9.0cm,height=7.0cm]{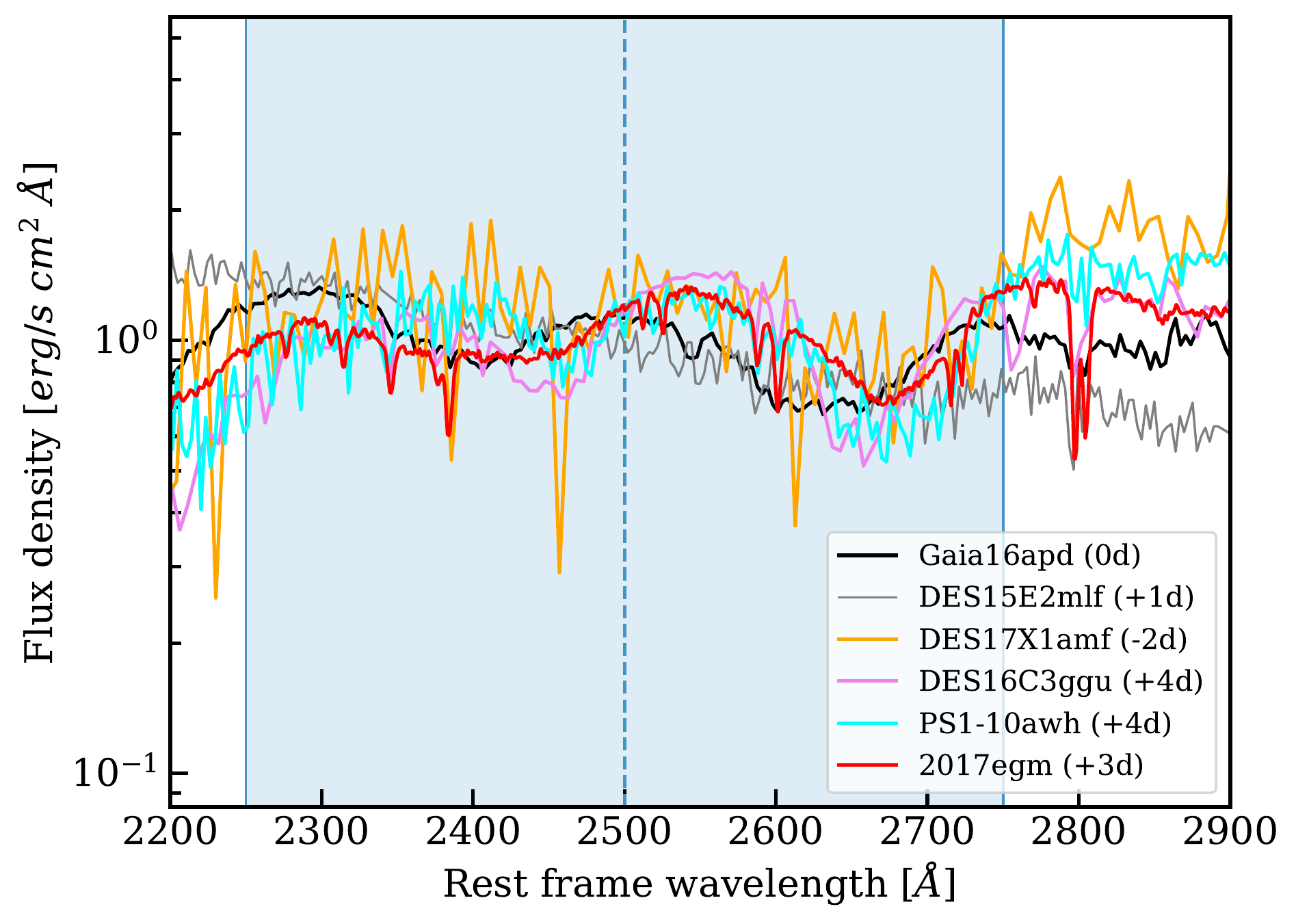}
\caption{Normalised \emph{Gaia16apd} spectra at peak compared with available near peak spectra of 5 SLSNe from the SLSN-UV Test sample along with the  \SI{250}{\nm} synthetic band.}  
\label{fig:overlay}
\end{figure}

However, \SI{250}{\nm} filter contains moderate-strength features and therefore we use the \emph{Gaia16apd} spectrum at peak \citep{linyan2017} for calculating those magnitudes. We test this by comparing the 5 available UV near-peak spectra with the \emph{Gaia16apd} spectrum. Figure \ref{fig:overlay} shows normalised near-peak spectra of the 5 SLSNe along with the Gaia16apd peak spectrum and the 250nm synthetic band used in the analysis. Visually, we find that the existing data are sufficiently consistent with the assumption of a template spectra at this wavelength. Furthermore, we calculated the \textit{K}-corrections for these 5 objects using their spectra and found that the mean difference between them and those from the \emph{Gaia16apd} spectrum is about ~0.07 mag. As explained later in sec \ref{sec:risetime}, we add an error floor of 0.15 mag to the computed errors on peak absolute magnitude from GPR, to account for uncertainties in the K-correction. Hence, using \emph{Gaia16apd} spectrum as template enables us to have a statistically significant number of objects, provides a baseline for future comparison, and does not alter the results. We note that this work does not aim to make a precise cosmological measurement, but rather explores the possible use of high redshift SLSNe-I as standardisable candles in the UV wavelengths. We also use the \emph{Gaia16apd} spectrum for the \SI{190}{\nm} filter used in our higher redshift exploratory test.

For each SLSN, we \textit{K}-correct the estimated peak magnitude in an observer frame filter to one of the synthetic filters. The observer frame filter is chosen such that its central wavelength in the SLSN rest frame is closest to the synthetic band. We calculate the average integrated flux in the rest-frame filter (i.e., blue-shifted observed filter) and then compare it with the flux in the {\it Gaia16apd} spectrum (\SI{250}{\nm}) or blackbody function (\SI{310}{\nm}) synthetic band. The difference between the two is added as correction to the peak magnitude estimated with the GPR interpolation. As a result, the absolute peak magnitude in synthetic filters can be written as:
\begin{equation}
    M_0(250) = M_0(X) + K_{X \rightarrow 250}
\end{equation}
\begin{equation}
    M_0(310) = M_0(X) + K_{X \rightarrow 310}
\end{equation} 

\noindent where $M_0(X)$ is the peak magnitude in the chosen observer frame filter $X$ for SLSNe, such that $X$ after accounting for the cosmological redshift $(1+z)$ is closest to the target synthetic filter and $K_{X \rightarrow 250}$ is the \textit{K}-correction from observed band $X$ to the synthetic band.

\subsection{Bayesian Inference}
\label{sec:bayes}
We perform linear fits to the correlations studied here employing a Bayesian approach for a weighted linear regression using Markov Chain Monte Carlo (MCMC) sampling. Borrowing from Bayes' principles, this method provides posterior distributions of our correlation parameters and enables us to reflect on the uncertainties in our estimates. Additionally, this method is termed weighted because one can tailor the variance of the likelihood allowing for the uncertainties in both the $x$ and $y$ variables along with the intrinsic scatter. 

We use linear models to correlate the peak absolute magnitude with each of the light curve properties (Section \ref{sec:lcprop}). For a light curve parameter $x$, we model the peak absolute magnitude in a synthetic filter $\gamma$ (\SI{250}{\nm} or \SI{310}{\nm} band) as follows,
\begin{equation}
\label{eq:lin_model}
    M_{\gamma} = b_0 + b_1(x)
\end{equation}

\noindent where $b_0$ and $b_1$ are the model parameters. Bayes's theorem gives the posterior probability distribution of the model parameters as
\begin{equation}
\label{eq:bayes}
 P(\Theta | D) \propto P(D | \Theta) P(\Theta),
\end{equation}

\noindent where D is the vector for the observed SLSN light curve data and $\Theta$ denotes the vector for the model parameters ($b_0$, and $b_1$). For a sample of $N$ SLSNe, model parameters for each SLSN are marginalised over and the likelihood probability distribution $P(D|\Theta)$ can be written as
\begin{equation}
P(D | \Theta) = \prod_{i=1}^{N} P(D_i | \Theta)    
\end{equation}

\noindent where $i$ is the index for the $N$ SLSNe of the sample (here $N$=13 for the key SLSN-UV sample) and $D_i$ is the observed light curve data for the $i$th SLSN. Assuming normally distributed errors and treating the peak absolute magnitude ($M_{\gamma}$) as the target variable, the log likelihood can be written as

\begin{equation}
    \ln \mathcal{L} = -\frac{1}{2}\sum_{i=1}^{N} \frac{(M_{\gamma}^i-M_{\gamma}^T)^2} {\sigma_{i}^2}  -\frac{1}{2} \sum_{i=1}^{N} \ln 2 \pi \sigma
    _{i}^2
\end{equation}

\noindent where $M_{\gamma}^i$ is the measured peak absolute magnitude of $i$th SLSN in the synthetic filters $\gamma$ and $M_{\gamma}^T$ is the true magnitude given by the model in equation \ref{eq:lin_model}. The variance $\sigma_i$ for a SLSN is computed as the quadrature sum of the errors on light curve data. We also include an additional intrinsic scatter term $\sigma_{int}$. This term is added to the variance and is left as a free parameter in the analysis accounting for any \emph{"unexplained"} dispersion observed in the peak absolute magnitudes. The errors on peak magnitudes are estimated from the GPR fits and the error on rise time are calculated by resampling as explained in section \ref{sec:timeerror}. The variance of the likelihood is then given as
\begin{equation}
    \sigma_i^2 = \sigma^2_{M_{\gamma, i}} + (b_1 \sigma_{\tau_{\mathrm{rise}}, i})^2 + \sigma_{int}^2
\end{equation}

The term $P(\Theta)$ in equation \ref{eq:bayes} are the priors on the model parameters. We adopt normal priors for the correlation parameters and a Half Cauchy distribution for the intrinsic scatter. The MCMC sampling is implemented using the "No U-Turn Sampler" (NUTS) provided in the \texttt{PyMC3}\footnote{See:\href{https://docs.pymc.io/}{https://docs.pymc.io/}} \citep{pymc3}, a \texttt{python} probabilistic programming package, with $10^5$ iterations. For all linear fits performed in this work, we use the observed light curve data parameters as input and estimate the posterior distributions for the correlation parameters $b_0$ and $b_1$, along with the intrinsic scatter. All best-fit values provided in this work are the posterior means and the errors in the parameters are the standard deviation of their posterior.


\section{Results}
\label{sec:results}
We measure the peak magnitudes and the light curve evolution properties using the GPR interpolated light curves for the literature sample. We proceed to investigate their peak magnitude correlations modelled using Bayesian regression. We note that all correlations are measured using only the SLSNe in the SLSN-UV test sample

\subsection{Peak Magnitude distribution}
\label{sec:peak_mag_dist}

\begin{figure}
\centering
\hspace*{-0.5cm}
\includegraphics[width=8.0cm,height=8.0cm]{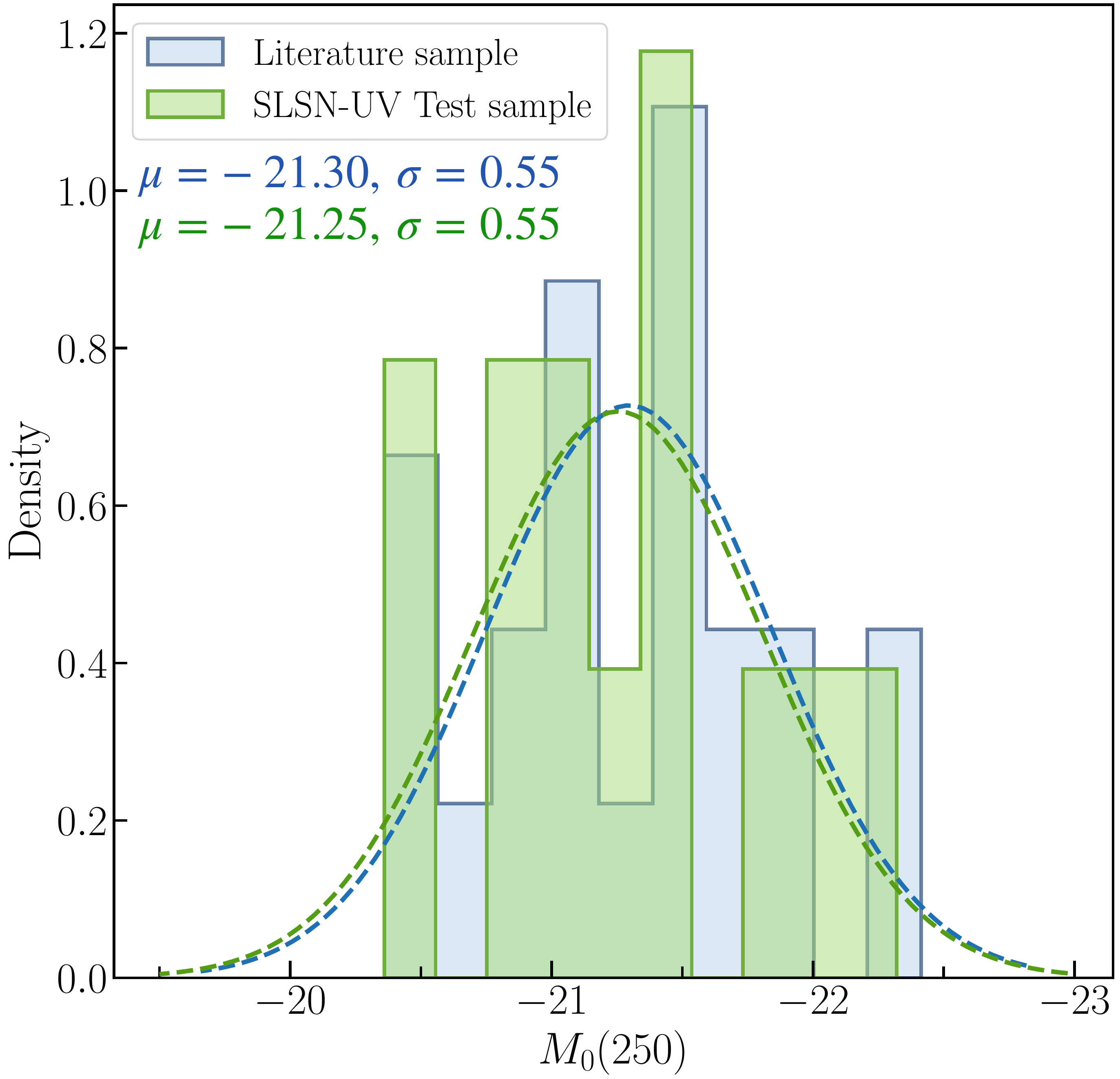}
\caption{Uncorrected peak absolute magnitude ($M_0(250)$) distributions of the full Literature Sample (22 objects) and the SLSN-UV test sample (13 objects). Normal distributions with the respective means and standard deviations are plotted in dashed lines.}
\label{fig:mag_dist}
\end{figure}

As a first check, we plot the distributions of \textit{uncorrected} peak magnitudes $M_0(250)$ for our Literature Sample and its subset, the SLSN-UV test sample. Figure~\ref{fig:mag_dist} shows the $M_0(250)$ histograms and density distributions from the GP interpolated light curves. The Literature Sample has an absolute magnitude mean of $-21.30$ with a standard deviation of $0.55$ and the SLSN-UV test sample has very similar values with a mean of $-21.25$ and standard deviation of $0.55$. We also measure the mean and spread for the bumpy (and may-be bumpy) population in our sample. These 10 SLSNe from the Literature Sample have a mean absolute magnitude of $-21.25$ with standard deviation of $0.54$, and 7 of these 10 included in the SLSN-UV test sample have a mean of $-21.39$ and standard deviation of $0.51$. These suggest that the various sub-samples are a good representation of the whole literature set and their selection has not introduced any arbitrary biases.

The scatter in the uncorrected \SI{250}{\nm} band peak magnitude distributions for the full Literature Sample as well as the SLSN-UV test sample are higher than those measured by \citetalias{inserra2014} in the \SI{400}{\nm} band. Differences between these two works include the rest-frame filters probed, redshift range of the samples (\citetalias{inserra2014} has $0.1 < z < 1.2$), and the sample selection criteria. Notably, \cite{lunnan2018} in the PS1 SLSN sample looked at peak magnitude distribution at \SI{260}{\nm} rest frame and found a even higher scatter of 1.15.

\subsection{Peak magnitude - Rise time relation}
\label{sec:risetime}
The maximum brightness of a SLSN has been shown to be dependent on the shape of its optical light curve or, more specifically, the rate of decline of the optical light curve. This correlation reduces the scatter observed in the \textit{uncorrected} peak magnitude distributions and has been used to demonstrate that SLSNe have the potential to be standardised for measuring cosmological distances \citepalias{inserra2014,inserra2020}. As described in Section~\ref{sec:peak_mag_dist}, the scatter in the raw \SI{250}{\nm} band peak absolute magnitudes is $\sim$0.55 mag. Here we investigate whether this scatter decreases using the correlation amomg the peak magnitude and the rise time in the rest-frame UV.

Previous works on SLSNe-I standardisation \citepalias[e.g.,][]{inserra2014, inserra2020} use the declining part of the SLSN light curve to characterise its shape and correlate the peak magnitude in \SI{400}{\nm} band with the decline rate over different time scales (10, 20, and 30 days). In this work, we choose instead to study the rising behaviour of the SLSN, characterised by rise time  $\tau_{\mathrm{rise}}^{\Delta1 \mathrm{m}}$ (see Section \ref{sec:lcprop}), and determine the relationships in the UV bands (see Section \ref{sec:syn_bands}). Rise time here is defined as the time elapsed in the rest-frame as the SLSN rises from 1 magnitude below peak magnitude to the peak. In order to better characterise the rising trend of the light curve, it would have been useful also to measure the rise time earlier from
the peak, for instance 2 magnitudes before peak. However, the data at hand do not accommodate this investigation, since few objects (a statistically small number) have such early data.

The interpolated light curves are used to measure the peak absolute magnitudes and $\tau_{\mathrm{rise}}^{\Delta1 \mathrm{m}}$ for all the SLSNe. The errors on the peak absolute magnitude are estimated with GPR and are of the order $\sim 0.05$ mag.  As this error is small compared to the photometric errors, we add in quadrature an error floor of 0.15 mag to peak magnitude errors. This will help account for uncertainties from the \emph{K}-correction method adopted here \citep[e.g.,][]{smith2018}. The error floor is added in the variance of the likelihood to the peak magnitude error term. We use only the SLSN-UV test sample (Section \ref{sec:golddata}) to perform the final correlation fits. The fitting method employs a Bayesian approach as described in Section \ref{sec:bayes}.

\begin{figure}
\centering
\includegraphics[width=8.2cm,height=7.5cm]{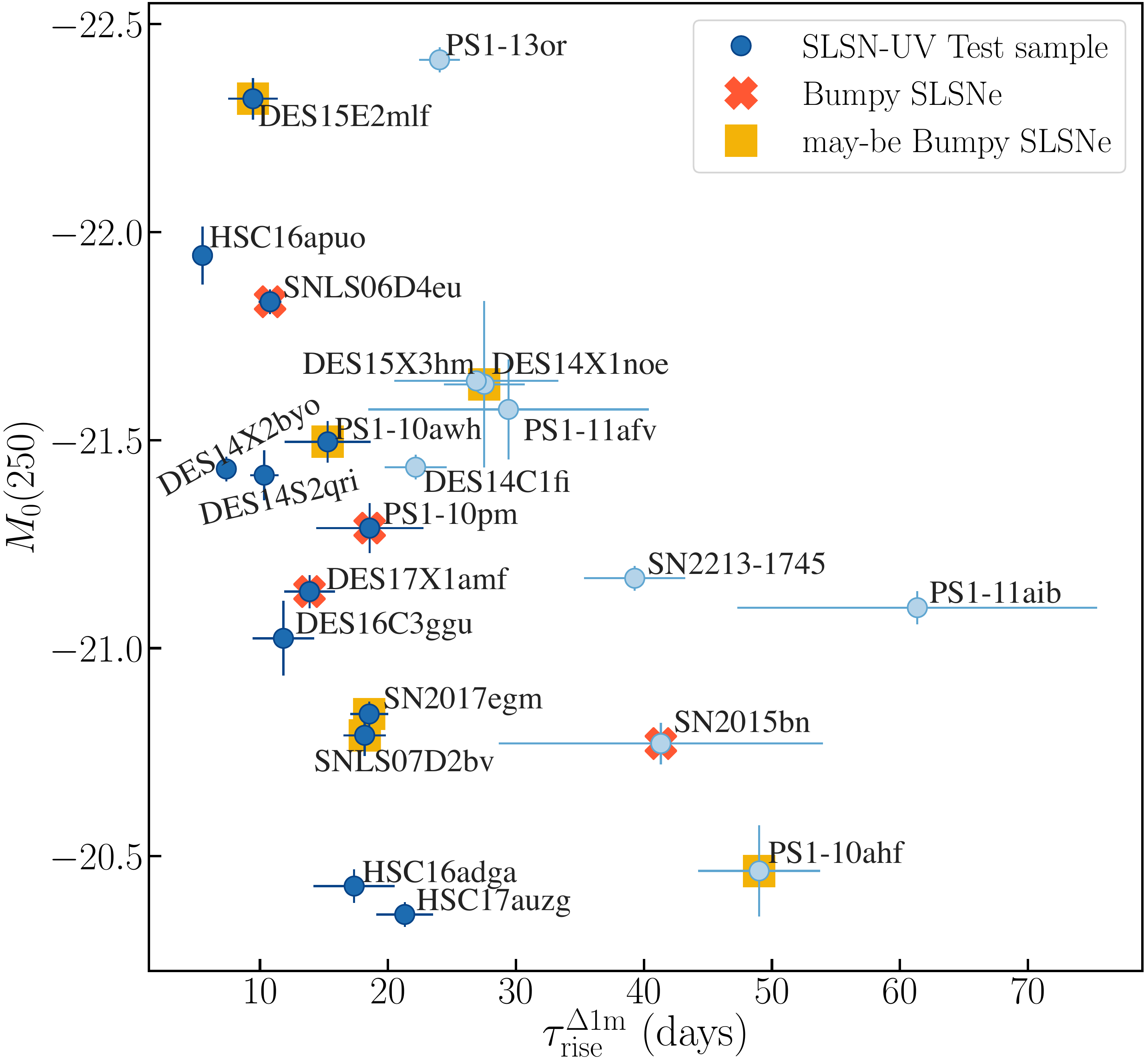}
\caption{Absolute peak magnitudes estimated in the \SI{250}{\nm} synthetic band versus rise time $\tau_{\mathrm{rise}}^{\Delta1 \mathrm{m}}$ for the literature sample. The SLSN-UV test sample, comprising 13 of the 22 SLSNe, are highlighted with dark blue circles. The bumpy and may-be bumpy SLSNe are marked with red crosses and yellow squares, respectively.}
\label{fig:Mrisetime}
\end{figure}

Figure~\ref{fig:Mrisetime} plots the peak absolute magnitude $M_0(250)$ versus the rise time $\tau_{\mathrm{rise}}^{\Delta1 \mathrm{m}}$ for the literature sample. The errors on both parameters are estimated as described in Section~\ref{sec:method}. Noting that the SLSN-UV test sample is a subset pulled from the literature sample, the SLSN-UV test sample are highlighted in dark blue while the remaining SLSNe are shown in light blue. SLSNe which show a bump (or possible bump) are marked in red circles (or yellow squares). 

The literature sample includes SLSNe with large peak epoch uncertainties, as seen in Figure~\ref{fig:Mrisetime}, and the events with the most complete data, i.e., the SLSN-UV test sample, are clustered in the short rise-time region of the plot. Focusing on this sample of 13 SLSNe, we observe a correlation where the brighter SLSNe in the \SI{250}{\nm} band rise faster. To quantify this relation and its scatter, we fit a linear function as in equation \ref{eq:lin_model} where the $x$ is $\tau_{\mathrm{rise}}^{\Delta1 \mathrm{m}}$ using Bayesian inference.

\begin{figure}
\centering
\includegraphics[width=8.0cm,height=7.3cm]{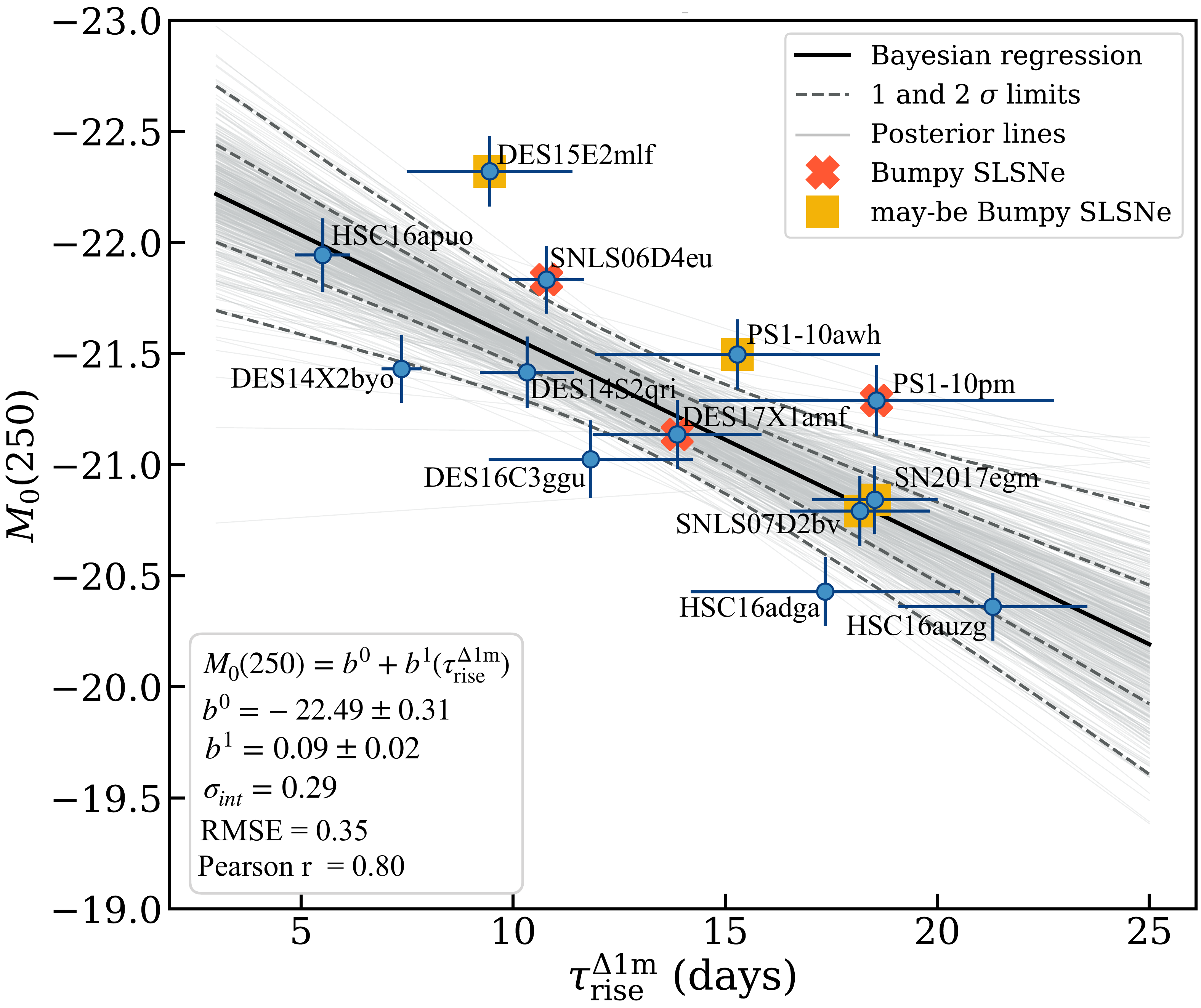}
\caption{
$M_0(250)$ vs.\ $\tau_{\mathrm{rise}}^{\Delta1 \mathrm{m}}$ correlation for the SLSN-UV test sample. The black solid line is the linear fit obtained using Bayesian regression. The light grey lines show the posterior fits with dashed lines representing the 1 and 2 $\sigma$ confidence intervals. The fit parameters, RMSE and Pearson's $r$ coefficient are given in the lower legend.}
\label{fig:corr_trise}
\end{figure}

The results are shown in Figure \ref{fig:corr_trise} that presents the linear fit to the SLSN-UV test sample, along with the posterior lines and sigma limits. The intrinsic scatter of the correlation is measured to be $0.29$ mag (roughly half the scatter for uncorrected magnitudes) with root mean square error (RMSE) of 0.35. The scatter in the \SI{250}{\nm} band rise time correlation is comparable to the scatter in the decline rate relationship in the \SI{400}{\nm} band measured by \cite{inserra2020}. The Pearson's $r$ coefficient is found to be $0.80$. 

\begin{figure}
\centering
\includegraphics[width=8.0cm,height=7.3cm]{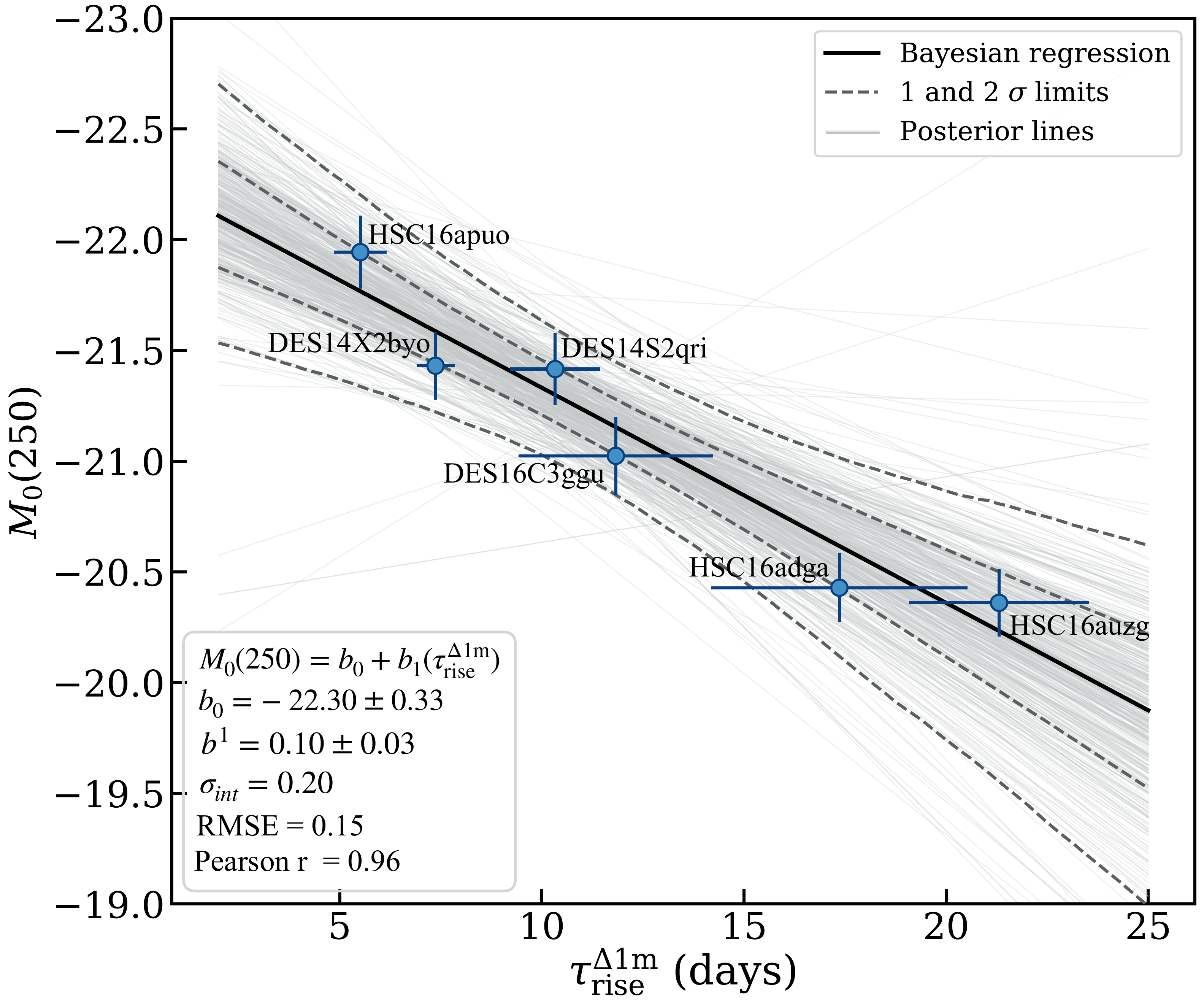}
\caption{Same as Figure~\ref{fig:corr_trise}, but for the SLSNe without a pre-peak bump among the SLSN-UV test sample.}
\label{fig:nobumpy250}
\end{figure}

\begin{figure}
\centering
\includegraphics[width=8.0cm,height=7.3cm]{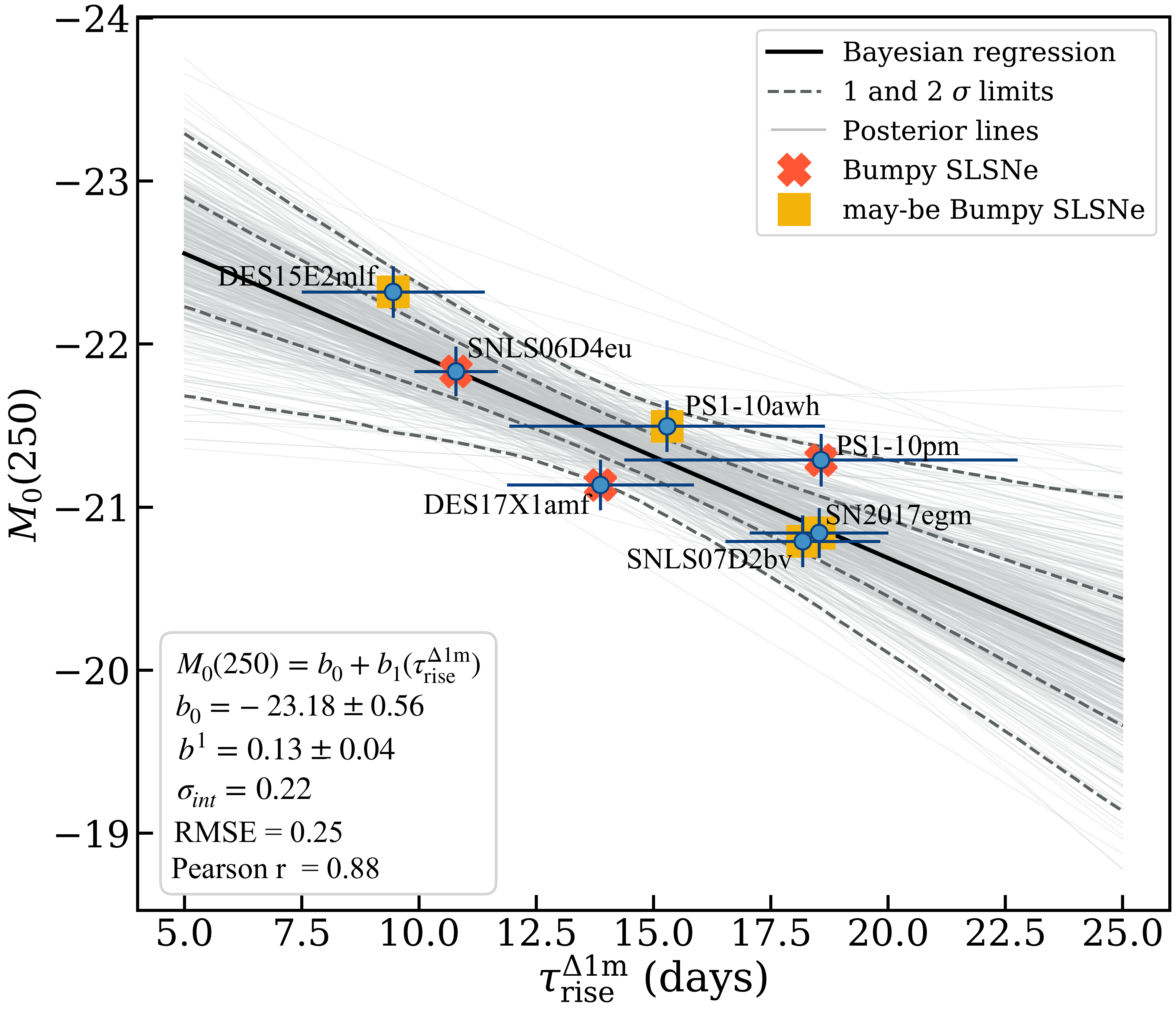}
\caption{Same as Figure~\ref{fig:corr_trise}, but for the bumpy (and may-be bumpy) SLSNe from the SLSN-UV test sample.}
\label{fig:bumpy250}
\end{figure}

Figure \ref{fig:corr_trise} also shows the 3 bumpy (red crosses) and 4 may-be bumpy (yellow squares) SLSNe that are present in the SLSN-UV test sample. These seven objects are observed to reside toward one side of the linear fit, possibly suggesting a similar relationship for bumpy objects but with longer rise times and/or brighter peak mags (i.e., they appear shifted in the x-axis or y-axis (or both) compared to the ones without a pre-peak bump), perhaps due to the added luminosity fade of the pre-peak bump. Under an assumption that bumpy SLSNe are a different population, potentially adding extra scatter in the relationship, we fit a linear relation excluding these 7 bumpy objects using our Bayesian framework as shown in Figure~\ref{fig:nobumpy250}. We see that while the correlation parameters ($b^0$ and $b^1$) remain similar (annotated in Figure \ref{fig:corr_trise} \&~\ref{fig:nobumpy250}), the intrinsic scatter reduces from 0.29 to 0.20 and the RMSE reduces from 0.35 to 0.15. This is a significant improvement if the underlying hypothesis is true, i.e., that bumpy SLSNe are a different population than those observed without a pre-peak bump. If so, these SLSNe can be eliminated from any cosmological sample based on their observed photometric behaviour. Finally, we fit a linear relation to the set of 7 Bumpy SLSNe data separately, as plotted in Figure~\ref{fig:bumpy250}. We find that the correlation parameters are similar to the two relations above with an intrinsic scatter of 0.22 and RMSE of 0.25, which is tighter than the values obtained with the full SLSN-UV test sample. 

As mentioned in Section \ref{sec:data}, the classification of three HSC SLSNe (HSC16apuo, HSCadga, HSCauzg) as type-I is unclear. We also evaluate the results excluding them that results in a linear fit scatter of 0.30 and a RMSE of 0.35. Hence, the fit results do not change significantly.

\subsection{Peak magnitude - Colour relation}
Past studies have shown SLSN peak magnitudes in rest-frame optical bands to be dependent on their colour at peak and on the rate of change of colour \citep{inserra2014}. We make measurements of the colour evolution of the SLSNe in our data sample and investigate their correlation with the peak absolute magnitude in the \SI{250}{\nm} band. Due to the paucity in data cadence of the SLSNe, the colour estimation is done using the GPR fitted curves that allows for a uniform method for measuring the colour and provides the errors on the values. Table \ref{tab:table2_slsnUV} lists the colour parameters for all the SLSNe in the Literature Sample.

Section \ref{sec:lcprop} gives the definition of \textit{colour at peak} and \textit{Delta colour}, a quantity to measure the change of colour between two epochs in a light curve. The colours here are computed as $M_d(250) - M_d(310)$ where $d$ is the epoch ($0$ for peak). The peak magnitudes in the \SI{310}{\nm} band for our sample are estimated in the same way as for the \SI{250}{\nm} band using the observed filter closest to \SI{310}{\nm} in the rest frame of the SLSN. The \emph{K}-corrections for the peak magnitude in \SI{310}{\nm} band ($M_0(310)$) are calculated using the \SI{15000}{\kelvin} blackbody curve. The colour at peak is then measured as $M_0(250) - M_0(310)$. We also calculate colour at 15 days before the peak $(d = -15)$ in a similar way. Furthermore, \textit{Delta colour} on the rising curve is then calculated as the difference between the colour at peak and colour at $-15$ days; $(250-310)_{-15} - (250-310)_0$, and is written as $\Delta(250-310)_{-15}$. All the objects in the literature sample have sufficient data to calculate the colours. However, there are two peculiar cases at high redshift ($z>2$) where the available filters were bluer than required. HSC16apuo ($z=3.22$), and SN2213-1745 ($z=2.05$) have effective rest-frame colour approximately as $185\nm - 212\nm$ and $205\nm - 252\nm$, respectively. We discuss the results below.

\begin{figure*}
\centering
\hspace*{-0.5cm}
\includegraphics[width=18.0cm,height=7.0cm]{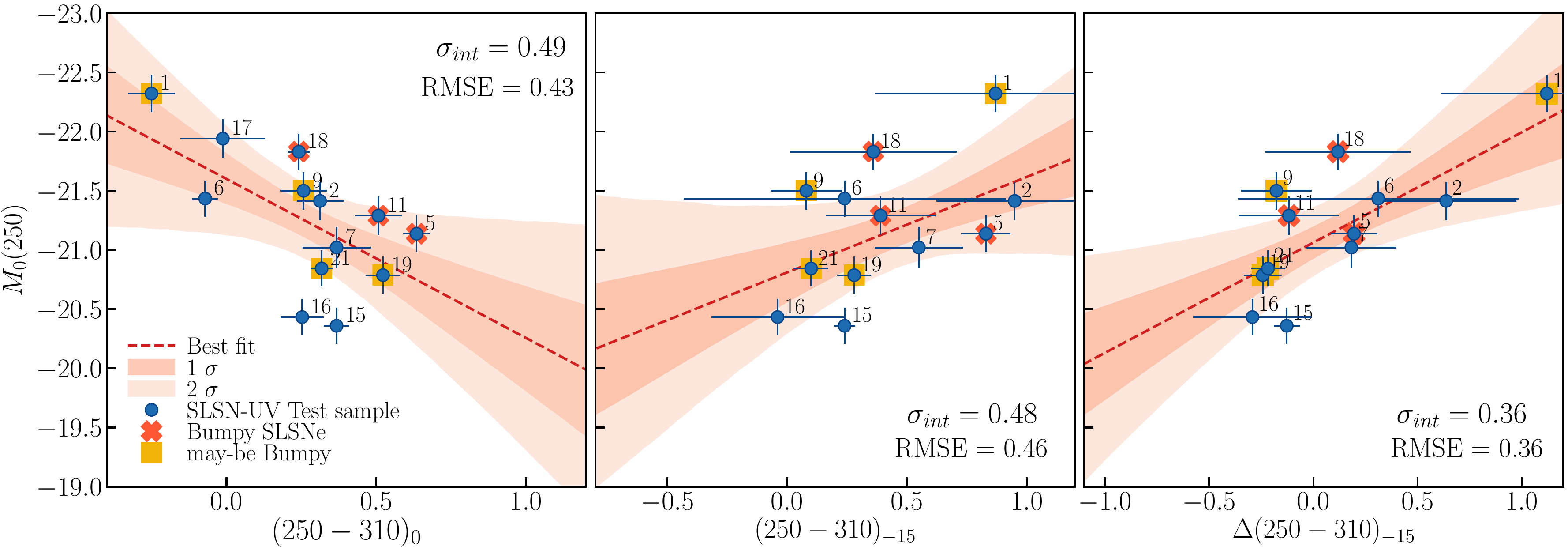}
\caption{Absolute peak magnitude $M_0(250)$ vs.\ colour at peak $(250-310)_0$ (first panel), with colour at 15 days before the peak $(250-310)_{-15}$ (second panel), and with change in colour over 15 days before the peak $\Delta(250-310)_{-15}$ (third panel) for the SLSN-UV Test sample. Dashed lines show the linear fit functions estimated using Bayesian regression and the shaded regions mark the 1 and 2 $\sigma$ confidence intervals. The bumpy (may-be bumpy) SLSNe in the sample are marked with red cross (yellow squares).}
\label{fig:colors}
\end{figure*}

Figure~\ref{fig:colors} plots the peak absolute magnitude $M_0(250)$ versus the colour at peak  $(250-310)_0$ (first panel), colour at 15 days before peak $(250-310)_{-15}$ (second panel), and \emph{Delta colour} $\Delta(250-310)_{-15}$ (last panel), for the 13 SLSNe in the key SLSN-UV sample. The plots also show the Bayesian linear fits to these objects along with their $1\sigma$ and $2\sigma$ confidence intervals. The intrinsic scatter and RMSE of all the three relations are annotated on their respective plots. The bumpy (may-be bumpy) SLSNe in the sample are marked with red cross (yellow squares).

We observe a weak correlation between peak magnitudes $M_0(250)$ and colour at peak, with a model intrinsic scatter of 0.46 mag and RMSE of 0.43. Fainter objects in the \SI{250}{\nm} filter appear to be redder at their peak. This is similar to the corresponding observation by \citetalias{inserra2014} in the \SI{400}{\nm} band. For the colour at 15 days before peak (second panel), the intrinsic scatter from fit is found to be 0.44 mag and RMSE of 0.64. The large errors in the colour data may contribute to the lower intrinsic scatter compared to the RMSE. The colour at 15 days before peak has larger errors because at early epochs many objects either have very sparse data or no data leading to inflated errors on the interpolated magnitudes. The correlation of peak magnitude with the change in colour over 15 days $\Delta(250-310)_{-15}$ is shown in the third panel. We find that during the rising phase, brighter objects tend to become redder faster. This fit has an intrinsic scatter of 0.31 mag and RMSE of 0.54. For the decline phase, \citetalias{inserra2014} found an opposite relationship; fainter objects become redder faster over 30 days after peak. We note that HCS16apuo (number 17 on the plots) is not included in the last two correlation fits because it does not have data 15 days before the peak, and hence, no reliable data points are available for the analysis.

\subsection{Rise rate and Decline rate}
Here we measure the rate of the SLSN light curve evolution during its rising and declining phases to study their relationship with the peak absolute magnitude. Rise rate ($\Delta M_{-15}$) is measured as change in the magnitude over 15 days before the maximum, and decline rates ($\Delta M_{15}$, $\Delta M_{30}$) are measured analogously over 15 or 30 days after the peak (see section \ref{sec:lcprop}). As described in section \ref{sec:risetime}, we observed a fairly good correlation ($\sigma_{int} = 0.29$) of peak magnitudes in UV ($M_0(250)$) with the rise time (\risetime) for the key SLSN-UV sample. Therefore a correlation with the rate of rise is an expected result. On the other hand, one of the most important \slsn cosmological correlations explored by \citetalias{inserra2014} and \citetalias{inserra2020} is the dependence of \SI{400}{\nm} band peak magnitude on the decline rate. We explore a similar relation here, but in the \SI{250}{\nm} band.

\begin{figure*}
\centering
\hspace*{-0.5cm}
\includegraphics[width=18.5cm,height=7.0cm]{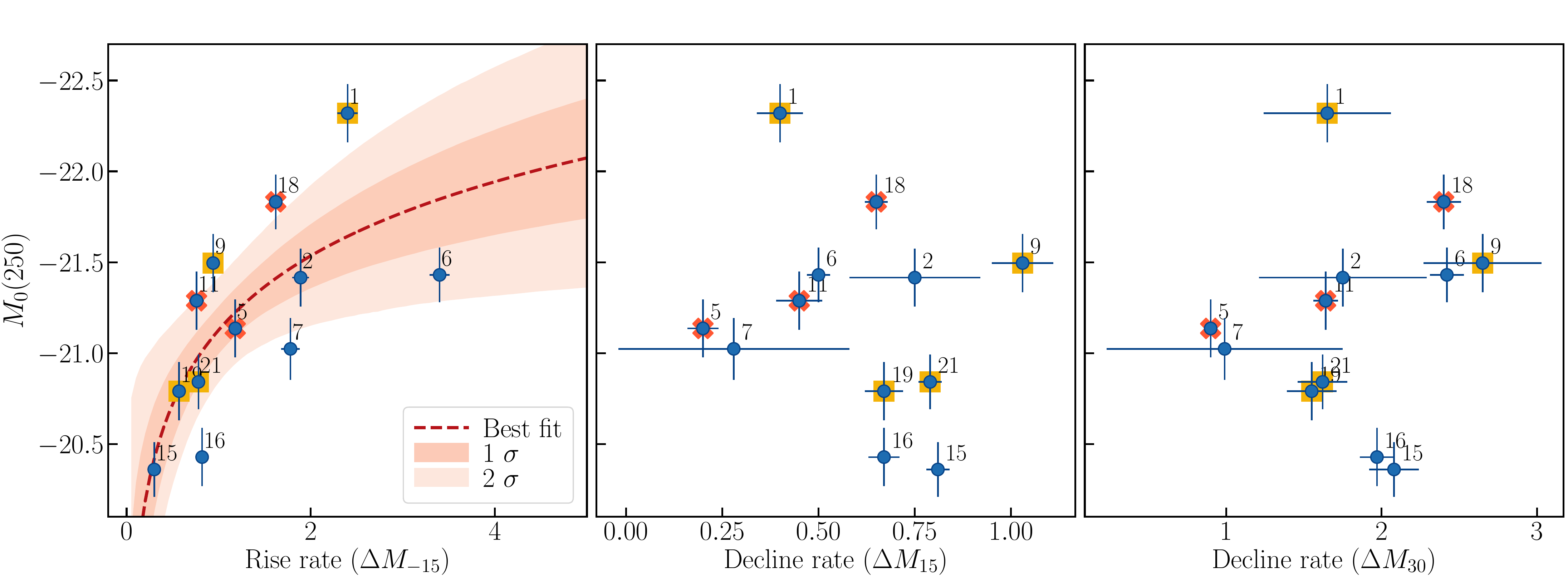}
\caption{Absolute peak magnitude $M_0(250)$ vs. rise rate over 15 days $\Delta M_{-15}$ (first panel), decline rate over 15 days $\Delta M_{15}$ (second panel), and decline rate over 30 days $\Delta M_{30}$ (last panel) for the SLSN-UV test sample. The dashed line in the left panel shows the best fit function estimated using Bayesian regression and the shaded regions mark the 1 and $2\sigma$ confidence intervals.} 
\label{fig:rates}
\end{figure*}

Figure \ref{fig:rates} plots the SLSN-UV test sample peak magnitude with the rise rate $\Delta M_{-15}$ (first panel), with the decline rate over 15 days $\Delta M_{15}$ (second panel), and lastly decline rate over 30 days $\Delta M_{30}$ (last panel). We observe a correlation between the rise rate and the peak absolute magnitude for the SLSN-UV test sample. Given that we observe a linear relationship between the rise time and the peak magnitude, and the rate of rise has dimension $1/\mathrm{time}$, we fit a function of the form $M = b^0 + b^1 \log (\Delta M_{15})$ instead of a linear fit. The best fit function estimated is shown in Figure \ref{fig:rates} along with  $1 \sigma$ and $2 \sigma$ confidence intervals. The intrinsic scatter of the relation is 0.41 mag. HSC16apuo (number 17) is not included in this fit, as it does not have sufficient early data as mentioned earlier. 

\begin{figure}
\centering
\includegraphics[width=7.0cm,height=7.0cm]{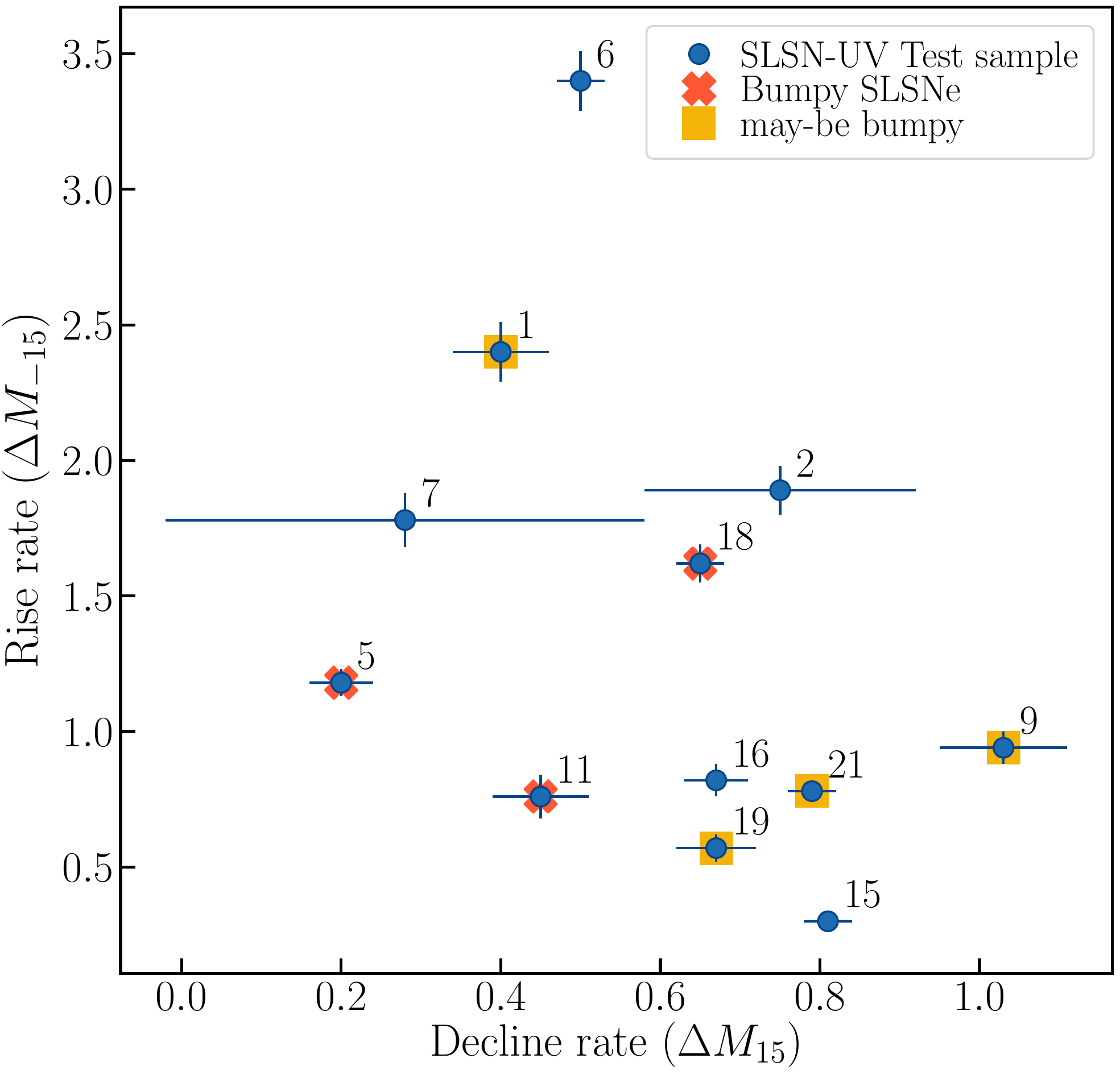}
\caption{Rise rate ($\Delta M_{-15}$) vs. Decline rate ($\Delta M_{15}$) plot for the SLSN-UV test sample. }
\label{fig:risevsdec}
\end{figure}

The second and third panel of Figure \ref{fig:rates} show the decline rate plots for 15 and and 30 days after maximum, respectively. For peak magnitude in UV ($M_0(250)$), we do not observe any correlation with the decline rate measured over 15 and 30 days. This is in contrast with the tight correlation ($\sigma = 0.33$) observed between peak magnitude and decline rate by \citetalias{inserra2014} and \citetalias{inserra2020} in the optical \SI{400}{\nm} band. In addition, our analysis also doesn't find correlation among the SLSNe-I rise and decline rates in the UV. Figure \ref{fig:risevsdec} shows rise rate vs decline rate for SLSN-UV test sample.

\subsection{Exploring the data for the highest redshift SLSNe-I}
Following our primary aim to investigate SLSN correlations in the rest frame UV, a sub-sample of the highest redshift SLSNe-I in our data enable an exploration of the rise time relation to wavelengths bluer than \SI{250}{\nm}. Studying the highest energy wavelength behaviour of SLSNe-I is vital to understand the physics behind these extreme events. A standardisation at wavelengths as close to Lyman-$\alpha$ as possible would enable the cosmological use of the highest redshift SLSNe-I.  

As mentioned earlier, one challenge in our investigation at wavelengths shorter than $\sim 3000 \A$ is the presence of strong broad absorption features. These features can potentially introduce scatter in any peak magnitude relation. In absence of a large sample of SLSN UV spectra, we make the same assumptions as we have for the \SI{250}{\nm} band, that all SLSNe-I exhibit similar spectral features near peak. Considering the SLSNe-I spectral library of \citep{quimby2018, linyan2018}, this assumption is reasonable and provides an important first step towards investigating high redshift SLSNe as cosmological probes in the FUV.

For our current SLSNe sample (in terms of their redshift and available data), the bluest regions which guarantee a statistically useful number of SLSNe cover a synthetic filter centred at \SI{190}{\nm} (see Section~\ref{sec:syn_bands} for details). Six SLSNe in the literature SLSN-UV sample have photometric data coverage close to 190 nm in their
rest frame, and their data quality in the bluest filter passes
the quality criteria defined in Section \ref{sec:golddata}. Similar to other synthetic bands, we fit light curves in the observed filters that are closest to \SI{190}{\nm} in the supernova rest-frame and employ GPR to estimate the peak absolute magnitude $M_0(190)$ and the rise time \risetime. The error on rise time is calculated by Monte Carlo resampling (section \ref{sec:timeerror}). We calculate cross-filter $K$-corrections for the peak magnitudes using the \textit{Gaia16apd} spectrum as a standard. Any results obtained here can be scaled and corrected to a better standard in the future with a larger spectral sample. 

\begin{figure}
\centering
\hspace*{-0.5cm}
\includegraphics[width=8.0cm,height=8.0cm]{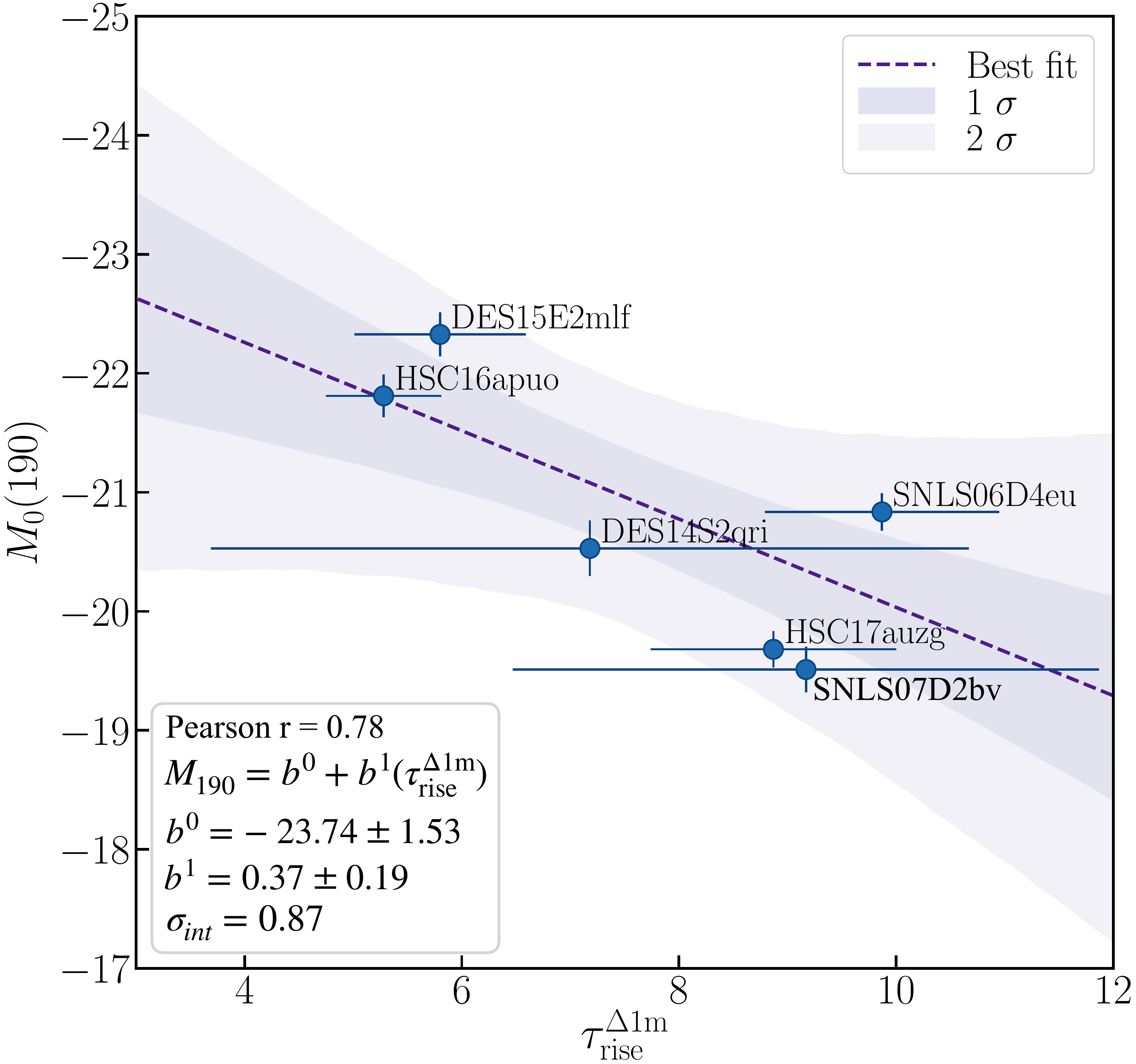}
\caption{$M_0(190)$ versus \risetime\ for $z>1.5$ SLSNe with sufficiently blue photometry in the rest-frame. The dashed line is the linear fit to the data obtained using Bayesian regression and the shaded regions show the $1 \sigma$ and $2 \sigma$ confidence intervals. The fit parameters and Pearson's $r$ coefficient are given in the legend.}
\label{fig:corr_1900}
\end{figure}

Figure \ref{fig:corr_1900} shows the peak magnitude in \SI{190}{\nm} filter $M_0(190$) versus the rise time (\risetime) for these six SLSNe-I. Supernovae have been observed to typically evolve faster at bluer wavelengths, and the \SI{190}{\nm} rise times of about $\sim$5--10 days confirm this expectation when compared to the redder bands, i.e., \SI{250}{\nm}  where the rise times are longer ($\sim$5--20 days).

The light curve peak absolute magnitude is observed to be correlated with rise times, however, with a steeper slope as compared to \SI{250}{\nm}. The fit function obtained with the Bayesian regression analysis is shown as the dashed line in Figure \ref{fig:corr_1900} along with $1 \sigma$ and $2 \sigma$ confidence interval limits. The intrinsic scatter of the fit is $0.87$ mag and the posteriors of the correlation coefficients are very sensitive to the prior information. We provide normal priors motivated by the values of the correlation parameters estimated with the least square fitting method. The slope of the relation is $0.37 \pm 0.19$, a steeper value as compared to the $0.09 \pm 0.02$ estimated for the $M_0(250)$ - \risetime relation. This result, along with the other results presented here, albeit with small data sets, motivate further exploration of SLSNe as cosmological tools at high redshift.

\subsection{Are Bumpies different?}\label{sec:different}

\cite{Nicoll2016} first suggested that pre-peak bumps may be ubiquitous in SLSNe light curves and may not be different populations. That is, the pre-peak bumps for those without evidence in their light curves may be the result of insufficient depth and/or lack of early data. However, there has not been any conclusive study on this topic to date. \cite{angus2019} confirmed existence of SLSNe without any pre-peak bump within DES data. We assess the peak magnitude vs.\ rise time correlation in \SI{250}{\nm} band  (Section~\ref{sec:risetime}) where we separately mark the bumpy SLSNe-I in the SLSN-UV test sample. In Figure \ref{fig:corr_trise}, bumpy (and may-be-bumpy) SLSNe show a similar peak magnitude vs.\ rise time relationship as the non-bumpy objects in the \SI{250}{\nm} filter. However, we note an apparent offset between the bumpy and non-bumpy $M_0(250)$--\risetime relationships, where the former appear slightly shifted to the upper right suggesting that they are either brighter, slower evolving, or both. This is indicated in the measured scatters when the correlation is determined separately with each of them, as shown in Figures \ref{fig:nobumpy250} and \ref{fig:bumpy250}. No clear trends are found in correlations with other parameters, such as colour or decline rate (Figures \ref{fig:colors} and \ref{fig:rates}). A more luminous peak and/or a slower rise time can be attributed to an addition of a separate pre-peak bump when convolved with the main burst light curve, if indeed the main bursts follow a standard relation. 

\begin{figure}
\centering
\includegraphics[width=8.0cm,height=8.0cm]{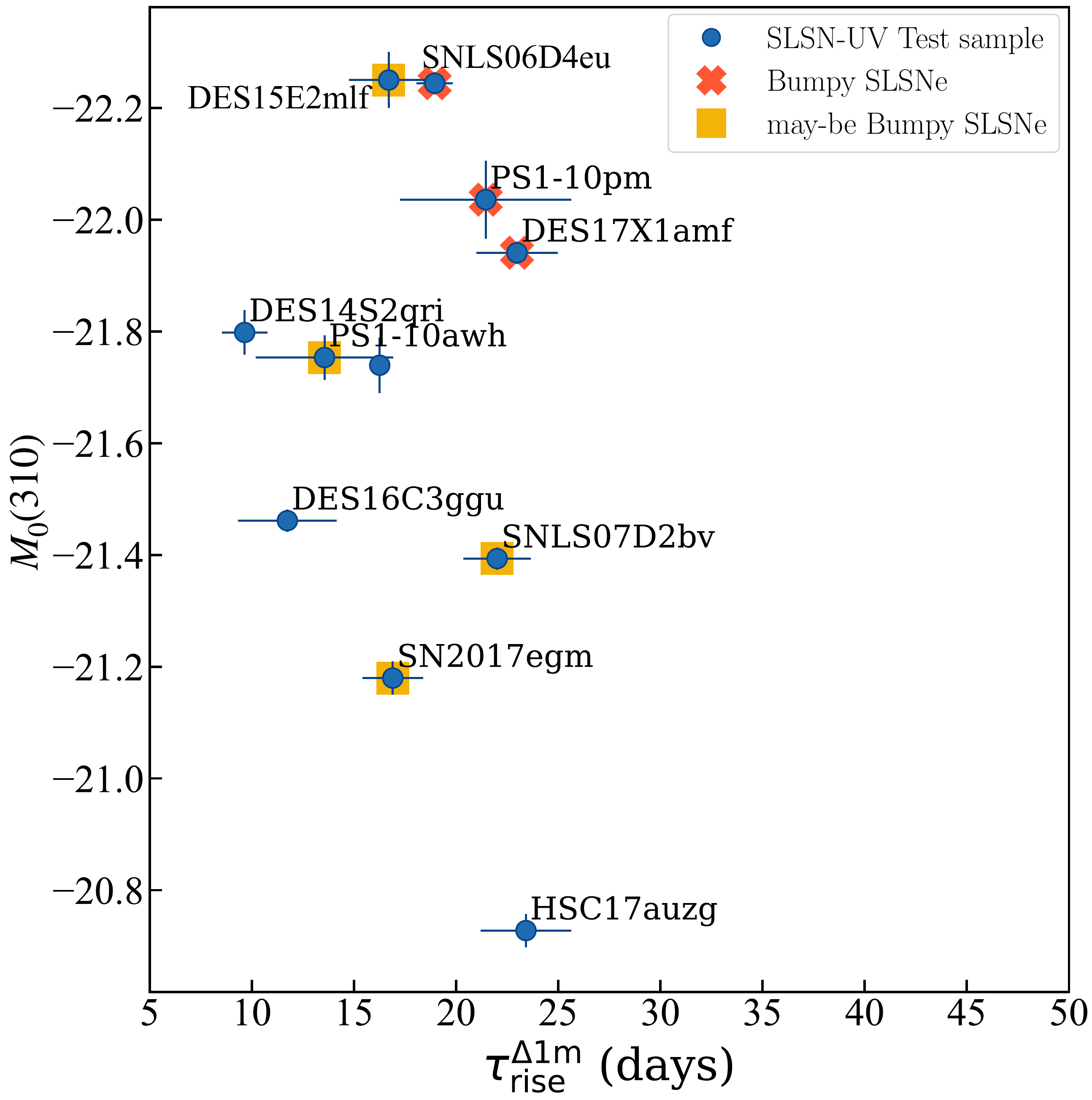}
\caption{$M_0(310)$ versus \risetime also measured in \SI{310}{\nm} band for SLSN-UV test sample. The bumpy (may-be bumpy) SLSNe in the sample are marked with red cross (yellow squares)}
\label{fig:3100}
\end{figure}

To explore the populations further, we measure the peak absolute magnitude and rise times in the \SI{310}{\nm} synthetic band for the SLSN-UV test sample, owing to the data availability (Figure \ref{fig:3100}). Two objects namely HSCadga and HSCapuo with redshifts 2.4 and 3.2 respectively are not included here because they lack data in rest-frame \SI{310}{\nm} filter. The bumpy (may-be bumpy) SLSNe are marked with red cross (yellow squares). In Figure \ref{fig:3100} SLSNe-I identified having a confirmed pre-peak bump are shifted to the upper right similar to what was observed for the \SI{250}{\nm} correlation although the may-be-bumpy SLSNe do not stand out. Additionally, the whole population has a relatively broader distribution compared to $M_0(250)$. Given the very limited data set, one can not draw a strong conclusion on whether or not pre-peak bumps in SLSN-I are ubiquitous, but the present analysis suggests that they could be different sub-class.

A KS test on the two samples gives a p-value $>> 0.05$ indicating that we can not reject that both SLSN-UV and the bumpy sample SLSNe are drawn from the same distribution. However, we do observe a significant reduction in scatter in peak magnitude-rise time correlation when Bumpy SLSNe are treated separately suggesting two distinct sub-populations. Larger samples by future surveys with deep photometric sensitivities and early and consistent cadence data are needed to properly test this hypothesis.

\section{Summary and Perspectives}
\label{sec:conclusion}
High redshift SLSNe are, and will be detected by their rest-frame UV emission with current and future optical/NIR surveys. This work presents a preliminary attempt to explore the rest-frame UV of SLSNe type I in context of their use as cosmological probes at high redshift. If standardisable in the UV, SLSNe would provide an effective tool for measuring cosmological parameters from $z\sim$ 0--20. SLSNe would provide a complementary method to SNe Ia to measure the Hubble diagram up to $z\sim1.2$ as well as a means to extend it beyond that, well into the epoch of deceleration, enabling us to distinguish between various dark energy models. 

The data sample compiled for the work is a set of 22 published SLSNe-I in redshift range $z\sim$ 1--3, referred to as the literature sample. We apply data quality cuts to the literature sample in order to distinguish objects with the best and/or most complete available UV/NUV data, and this sub-set of 13 SLSNe-I is called the SLSN-UV test sample. All the cosmological correlations are determined using the SLSN-UV test sample. We also identify SLSNe which have, or most likely have, an early pre-peak bump in their light curve, and refer to them as bumpy SLSNe. With the aim of exploring rest-frame UV, and given the redshift range of the sample and available data, we chose a synthetic filter centered at \SI{250}{\nm} to analyse the peak magnitude correlations. All the SLSNe light curves are fit using GPR to avoid any assumption on the light curve shape (see section \ref{sec:lcfitting}), and these GP interpolated light curves are used to estimate the peak absolute magnitudes ($M_0(250)$) and the light curve properties (along with their uncertainties). We apply cross-filter \textit{K}-corrections to the estimated peak magnitudes from the observer frame filter into the \SI{250}{\nm} band using the {\it Gaia16apd} UV spectrum at peak, and into the \SI{310}{\nm} band (used for calculating colors) with a \SI{15000}{\kelvin} blackbody function assuming them as standards. Peak magnitude correlations are modelled using a Bayesian framework which constraints the posteriors of the correlation coefficients along with the intrinsic scatter in the relation. The main results of the work are summarised in the following.

\begin{itemize}
    \item We observe a linear correlation between $M_0(250)$ and \risetime for the SLSN-UV test sample where brighter objects appear to rise faster. The intrinsic scatter of the model, $\sigma_{int}$, is found to be 0.29 mag (RMSE = 0.35). If we exclude those SLSNe which are identified as having pre-peak bumps (bumpy SLSNe-I), the $M_0(250)$ -- \risetime correlation becomes tighter with $\sigma_{int} = 0.2$ and RMSE = 0.15. With or without the bumpy SLSNe-I, this result strongly encourage further investigations into their use as high redshift cosmological probes in the rest-frame UV.
    
    \item For the colour evolution of SLSN light curves, we correlate the peak magnitude with three quantities; colour at peak, colour at 15 days before peak, and the rate of change of colour between these two epochs. The colour terms are calculated using \SI{250}{\nm} and \SI{310}{\nm} magnitudes. We observe correlations, albeit weak, for the three color quantities with the peak magnitude $M_0(250)$ for the SLSN-UV test sample. Relatively stronger relationship is seen between $M_0(250)$ and delta color $\Delta (250-310)_{-15}$ with a scatter of 0.37. 
    
    \item Peak magnitude versus rise rate relation ($\Delta M_{-15}$) reproduces the result similar to that observed for the rise time. More interestingly though, we do not observe any correlation of peak magnitudes in \SI{250}{\nm} band with the decline rate over 15 and and 30 days ($\Delta M_{15}$ and $\Delta M_{30}$). This is contrary to what has been observed for peak magnitude in \SI{400}{\nm} band \citetalias{inserra2014,inserra2020}.
    
    \item Six high redshift SLSNe from the SLSN-UV test sample have photometric data that enable an analysis at a synthetic band centred at \SI{190}{\nm} in their rest-frame. We explore this band in order to investigate the peak magnitude versus rise time relation at bluer wavelengths for using SLSNe at very high redshifts. We observe  a correlation with an intrinsic scatter $\sigma_{int}$ = 0.87 (RMSE = 0.67). 
 
    \item We also perform peak magnitude-rise time correlation in the \SI{310}{\nm} band for the test sample SLSNe which have the required data. This is done to examine whether Bumpy SLSNe show any distinguishing trend in this band similar to the \SI{250}{\nm}. We find that the three SLSNe with confirmed pre-peak bumps in their light curve are shifted to the upper right corner of the plot similar to what was observed for the \SI{250}{\nm} correlation. This indicates that pre-peak bumps might not be a universal phenomena among SLSNe-I, however, it is difficult to draw a conclusive result given the small data set. Additionally, comparing the rise times at three different wavelength bands, we observe that \risetime show a consistent decrease with decreasing wavelength, with rise times of $\sim$10--30 days at 310 nm, $\sim$10--20 days at 250 nm, and $\sim$5--10 days at 190 nm.
\end{itemize}

The results obtained in this work are highly promising for the use of SLSNe-I as high redshift cosmological probes in their rest-frame UV. We have adopted an unbiased approach for data selection, using all available data with appropriate photometry. However, the data set is yet small for a very robust analysis. Larger samples in future are strongly encouraged to test the observed correlations. Furthermore, SLSNe-I UV spectroscopy is critical to test for absorption-line consistency at wavelengths from 1216 (Lyman-$\alpha$) to $\sim$3000 \AA, where SLSNe-I continua show absorption. 

We note that if the Bumpy and non-Bumpy populations are indeed different, the peak magnitude vs.\ rise time relationship is significantly improved when eliminating Bumpy (and may-be-bumpy) SLSNe-I and is almost comparable to SNe Ia, as demonstrated in Section~\ref{sec:risetime}, with the important caveat of a small sample. With sufficiently deep photometry, Bumpy SLSNe-I can be identified by their distinct light curves and removed from cosmological samples. Additionally, we would like to highlight an interesting inference that at UV wavelengths, SLSNe peak magnitudes exhibit a relatively tighter correlation with the rising phase of the light curve (here \risetime) instead of the decline which is the common parameter used in literature for SLSNe relations in optical bands. With the present sample, we do not observe any trend for $M_0(250)$ with the decline rate. Due to data limitations, we can not perform this analysis in the \SI{400}{\nm} band for a direct comparison with \citetalias{inserra2014,inserra2020}. It would be natural to explore rise time correlations at optical wavelengths with a larger sample in future.

Finally, a key challenge for surveys that aim to detect very high redshift SLSNe is the very long (1 $+z$) time-dilated search baselines needed to detect their evolution. For example, for an assumed rest-frame 50-day rise and overall 200-day evolution at z $\sim$ 15 becomes $\sim$2 and $\sim$9 years, respectively.  Supernovae are observed to typically evolve faster at shorter wavelengths and this work confirms this relation to the NUV and FUV, with rise timescales increasingly shorter, and as short as $\sim$5--10 days for rise from 1 magnitude to peak at wavelengths near $\sim$1900\AA.  The comparatively shorter evolution timescales for SLSNe-I in the UV compared to the optical will help mitigate the expected extremely long duration evolution of SLSNe at the highest redshifts (z $\sim$ 6--20), making surveys more practical and enabling their detection in surveys with cadences designed for other types of lower redshift supernovae.  

Surveys for high redshift SLSNe require both deep imaging capability (m $\gtrsim$ 26 per filter, per epoch) and wide fields to discover these relatively rare events.  However, their high utility outlined in Section~\ref{sec:intro}, including galaxy and stellar evolution ISM, CGM, and IGM probes, detection of the deaths of Population III stars, searching for pair-instability events, and cosmological probes warrant such surveys.  SLSNe can be detected to $z\sim6$ using optical facilities such as CTIO Dark Energy Camera, NAOJ Subaru Hyper-SuprimeCam, and the future Keck Wide-Field Imager.  SLSNe to $z\sim20$ can be detected by facilities such as the Nancy Roman Space Telescope, Euclid, the University of Tokyo Atacama Observatory SWIMS, and the Kunlun Dark Universe Survey Telescope (KDUST).

\section*{Acknowledgements}

NK is supported by a grant from VILLUM FONDEN (project number 16599). NK would like to thank Charlotte Angus, Mat Smith, Luca Izzo for the very helpful discussions. We also thank L.Yan for providing us the reduced spectra for Gaia16apd.  Part of this research was funded by the Australian Research Council Centre of Excellence for Gravitational Wave Discovery (OzGrav), CE170100004 and the Australian Research Council Centre of Excellence for All-sky Astrophysics in 3 Dimensions (ASTRO-3D), CE170100013. MB acknowledges financial support from MIUR (PRIN 2017 grant 20179ZF5KS).


\section*{Data Availability}

All data underlying this article is available within the article and enlisted in Table \ref{tab:table1} and Table \ref{tab:table2_slsnUV}. In addition, individual SLSNe-I light curves and spectra analysed in this article are available to the public via their corresponding papers which are referenced in Table \ref{tab:table1}.

\bibliographystyle{mnras}
\bibliography{ref} 




\bsp	
\label{lastpage}
\end{document}

%% file: tables/Table1_redo.tex
\begingroup
\setlength{\tabcolsep}{8pt} 
\renewcommand{\arraystretch}{1.2} 
\begin{tabular}{l c c c c c c c}
\hline\hline
ID & $z$ & Filters used for & Rest frame coverage  & Rest frame coverage & SLSN-UV test & Bumpy & Literature\\
 &  & 250, 310 nm & for \SI{250}{\nm} (\A)  & for \SI{310}{\nm} (\A) & sample &  & reference\\
\hline
DES15E2mlf      &    1.860 & \emph{r} $\rightarrow$ 250, \emph{i} $\rightarrow$ 310 & $  1983 \rightarrow  2502 $ & $  2475 \rightarrow  2993 $ & Yes   & may-be bumpy & a \\
DES14S2qri      &    1.500 & \emph{r} $\rightarrow$ 250, \emph{i} $\rightarrow$ 310 & $  2268 \rightarrow  2862 $ & $  2832 \rightarrow  3424 $ & Yes   & No & a \\
DES14C1fi       &    1.302 & \emph{r} $\rightarrow$ 250, \emph{i} $\rightarrow$ 310 & $  2463 \rightarrow  3108 $ & $  3075 \rightarrow  3719 $ & No    & No   & a  \\
DES15X1noe      &    1.188 & \emph{r} $\rightarrow$ 250, \emph{i} $\rightarrow$ 310 & $  2592 \rightarrow  3270 $ & $  2592 \rightarrow  3270 $ & No    & may-be bumpy  & a   \\
DES17X1amf      &    0.920 & \emph{g} $\rightarrow$ 250, \emph{r} $\rightarrow$ 310 & $  2174 \rightarrow  2851 $ & $  2953 \rightarrow  3726 $ & Yes   & Bumpy  & a     \\
DES14X2byo      &    0.868 & \emph{g} $\rightarrow$ 250, \emph{r} $\rightarrow$ 310 & $  2235 \rightarrow  2930 $ & $  3036 \rightarrow  3830 $ & Yes   & No     &  a   \\
DES16C3ggu      &    0.949 & \emph{g} $\rightarrow$ 250, \emph{r} $\rightarrow$ 310 & $  2142 \rightarrow  2809 $ & $  2909 \rightarrow  3671 $ & Yes   & No      & a  \\
DES15X3hm       &    0.860 & \emph{g} $\rightarrow$ 250, \emph{r} $\rightarrow$ 310 & $  2244 \rightarrow  2943 $ & $  3049 \rightarrow  3847 $ & No    & No      & a    \\
PS1-10awh       &    0.908 & \emph{g} $\rightarrow$ 250, \emph{r} $\rightarrow$ 310 & $  2229 \rightarrow  2887 $ & $  2880 \rightarrow  3616 $ & Yes   & may-be bumpy  &  b \\
PS1-13or        &    1.520 & \emph{r} $\rightarrow$ 250, \emph{i} $\rightarrow$ 310 & $  2181 \rightarrow  2738 $ & $  2738 \rightarrow  3253 $ & No    & No      & c        \\
PS1-10pm        &    1.206 & \emph{r} $\rightarrow$ 250, \emph{i} $\rightarrow$ 310 & $  2491 \rightarrow  3128 $ & $  3128 \rightarrow  3716 $ & Yes   & Bumpy     & d      \\
PS1-11afv       &    1.407 & \emph{r} $\rightarrow$ 250, \emph{i} $\rightarrow$ 310 & $  2283 \rightarrow  2866 $ & $  2867 \rightarrow  3405 $ & No    & No      & c        \\
PS1-10ahf       &    1.100 & \emph{r} $\rightarrow$ 250, \emph{i} $\rightarrow$ 310 & $  2617 \rightarrow  3286 $ & $  2617 \rightarrow  3286 $ & No    & may-be bumpy  & d  \\
PS1-11aib       &    0.997 & \emph{g} $\rightarrow$ 250, \emph{r} $\rightarrow$ 310 & $  2129 \rightarrow  2758 $ & $  2752 \rightarrow  3455 $ & No    & No     & c         \\
HSC17auzg       &    1.965 & \emph{i} $\rightarrow$ 250, \emph{z} $\rightarrow$ 310 & $  2353 \rightarrow  2877 $ & $  2876 \rightarrow  3137 $ & Yes   & No    & e          \\
HSC16adga       &    2.399 & \emph{i} $\rightarrow$ 250, \emph{z} $\rightarrow$ 310 & $  2052 \rightarrow  2509 $ & $  2509 \rightarrow  2737 $ & Yes   & No      & e        \\
HSC16apuo       &    3.220 & \emph{z} $\rightarrow$ 250, \emph{z} $\rightarrow$ 310 & $  2021 \rightarrow  2204 $ & $  2021 \rightarrow  2204 $ & Yes   & No       & e       \\
SNLS06D4eu      &    1.588 & \emph{r} $\rightarrow$ 250, \emph{i} $\rightarrow$ 310 & $  2176 \rightarrow  2658 $ & $  2712 \rightarrow  3240 $ & Yes   & Bumpy     & f      \\
SNLS07D2bv      &    1.500 & \emph{r} $\rightarrow$ 250, \emph{i} $\rightarrow$ 310 & $  2252 \rightarrow  2752 $ & $  2808 \rightarrow  3354 $ & Yes   & may-be bumpy   & f   \\
SN2213-1745     &    2.046 & \emph{i} $\rightarrow$ 250, \emph{i} $\rightarrow$ 310 & $  2305 \rightarrow  2753 $ & $  2305 \rightarrow  2753 $ & No    & No      & g        \\
SN2017egm       &    0.030 & \emph{uvw1} $\rightarrow$ 250, \emph{u} $\rightarrow$ 310 & $  2187 \rightarrow  2860 $ & $  2989 \rightarrow  3745 $ & Yes   & may-be bumpy  & h  \\
SN2015bn        &    0.114 & \emph{uvw1} $\rightarrow$ 250, \emph{u} $\rightarrow$ 310 & $  2023 \rightarrow  2645 $ & $  2765 \rightarrow  3464 $ & No    & Bumpy      & i     \\
\hline
\end{tabular}
\endgroup

%% file: tables/slsnUV_table2.tex
\begingroup
\setlength{\tabcolsep}{8pt} 
\renewcommand{\arraystretch}{1.2} 
\begin{tabular}{c c c c c c c c c c c}
\hline\hline
ID & z & $M_0(250)$ & $\tau_{\mathrm{rise}}^{\Delta1 \mathrm{m}}$ & $\Delta M_{-15}$ & $\Delta M_{15}$ & $\Delta M_{30}$ & $(250-310)_0$ & $(250-310)_{-15}$ & $\Delta (250-310)_{-15}$ & Plotting \\
 &  &  &(days) & & & & & & & Number\\ 
\hline
DES15E2mlf      &    1.860 & $-22.32 \pm  0.05$ & $ 9.45 \pm  1.95$ & $ 2.40 \pm  0.11$ & $ 0.40 \pm  0.06$ & $ 1.65 \pm  0.41$ & $-0.25 \pm  0.08$ & $ 0.87 \pm  0.50$ & $ 1.01 \pm  0.51$ &     1 \\
DES14S2qri      &    1.500 & $-21.42 \pm  0.06$ & $10.33 \pm  1.11$ & $ 1.89 \pm  0.09$ & $ 0.75 \pm  0.17$ & $ 1.75 \pm  0.54$ & $ 0.31 \pm  0.08$ & $ 0.95 \pm  0.33$ & $ 0.68 \pm  0.34$ &     2 \\
DES14C1fi       &    1.302 & $-21.43 \pm  0.03$ & $22.16 \pm  2.42$ & $ 0.58 \pm  0.03$ & $ 0.37 \pm  0.03$ & $ 0.93 \pm  0.03$ & $ 0.32 \pm  0.04$ & $ 0.18 \pm  0.08$ & $ 0.06 \pm  0.09$ &     3 \\
DES15X1noe      &    1.188 & $-21.63 \pm  0.20$ & $27.52 \pm  3.16$ & $ 0.29 \pm  0.20$ & $ 0.22 \pm  0.23$ & $ 0.76 \pm  0.25$ & $ 0.39 \pm  0.26$ & $ 0.12 \pm  0.08$ & $ 0.02 \pm  0.28$ &     4 \\
DES17X1amf      &    0.920 & $-21.14 \pm  0.04$ & $13.87 \pm  1.99$ & $ 1.18 \pm  0.05$ & $ 0.20 \pm  0.04$ & $ 0.90 \pm  0.04$ & $ 0.63 \pm  0.04$ & $ 0.83 \pm  0.10$ & $ 0.19 \pm  0.11$ &     5 \\
DES14X2byo      &    0.868 & $-21.43 \pm  0.03$ & $ 7.37 \pm  0.47$ & $ 3.40 \pm  0.11$ & $ 0.50 \pm  0.03$ & $ 2.42 \pm  0.11$ & $-0.07 \pm  0.04$ & $ 0.24 \pm  0.67$ & $ 0.38 \pm  0.67$ &     6 \\
DES16C3ggu      &    0.949 & $-21.02 \pm  0.09$ & $11.83 \pm  2.41$ & $ 1.78 \pm  0.10$ & $ 0.28 \pm  0.30$ & $ 0.99 \pm  0.76$ & $ 0.37 \pm  0.11$ & $ 0.55 \pm  0.18$ & $ 0.13 \pm  0.22$ &     7 \\
DES15X3hm       &    0.860 & $-21.64 \pm  0.02$ & $26.90 \pm  6.40$ & $ 0.48 \pm  0.06$ & $ 0.62 \pm  0.02$ & $ 2.32 \pm  0.03$ & $ 0.13 \pm  0.02$ & $-0.38 \pm  0.28$ & $-0.43 \pm  0.28$ &     8 \\
PS1-10awh       &    0.908 & $-21.50 \pm  0.05$ & $15.29 \pm  3.36$ & $ 0.94 \pm  0.06$ & $ 1.03 \pm  0.08$ & $ 2.65 \pm  0.38$ & $ 0.26 \pm  0.08$ & $ 0.08 \pm  0.15$ & $-0.13 \pm  0.17$ &     9 \\
PS1-13or        &    1.520 & $-22.41 \pm  0.03$ & $24.03 \pm  1.60$ & $ 0.42 \pm  0.03$ & $ 0.50 \pm  0.05$ & $ 1.15 \pm  0.18$ & $ 0.06 \pm  0.04$ & $-0.23 \pm  0.08$ & $-0.32 \pm  0.09$ &    10 \\
PS1-10pm        &    1.206 & $-21.29 \pm  0.06$ & $18.57 \pm  4.19$ & $ 0.76 \pm  0.08$ & $ 0.45 \pm  0.06$ & $ 1.64 \pm  0.08$ & $ 0.51 \pm  0.08$ & $ 0.39 \pm  0.23$ & $ 0.15 \pm  0.24$ &    11 \\
PS1-11afv       &    1.407 & $-21.57 \pm  0.12$ & $29.41 \pm 10.96$ & $ 0.27 \pm  0.13$ & $ 0.06 \pm  0.17$ & $ 0.18 \pm  0.27$ & $ 0.44 \pm  0.39$ & $ 0.29 \pm  0.08$ & $-0.11 \pm  0.40$ &    12 \\
PS1-10ahf       &    1.100 & $-20.46 \pm  0.11$ & $49.00 \pm  4.78$ & $ 0.18 \pm  0.11$ & $ 0.19 \pm  0.14$ & $ 0.31 \pm  0.23$ & $ 0.59 \pm  0.14$ & $ 0.41 \pm  0.07$ & $ 0.13 \pm  0.16$ &    13 \\
PS1-11aib       &    0.997 & $-21.10 \pm  0.04$ & $61.36 \pm 14.06$ & $ 0.20 \pm  0.04$ & $ 0.22 \pm  0.05$ & $ 0.54 \pm  0.11$ & $ 0.67 \pm  0.05$ & $ 0.74 \pm  0.06$ & $-0.00 \pm  0.08$ &    14 \\
HSC17auzg       &    1.965 & $-20.36 \pm  0.03$ & $21.31 \pm  2.23$ & $ 0.30 \pm  0.03$ & $ 0.81 \pm  0.03$ & $ 2.08 \pm  0.16$ & $ 0.37 \pm  0.04$ & $ 0.24 \pm  0.04$ & $-0.07 \pm  0.06$ &    15 \\
HSC16adga       &    2.399 & $-20.43 \pm  0.04$ & $17.36 \pm  3.17$ & $ 0.82 \pm  0.06$ & $ 0.67 \pm  0.04$ & $ 1.97 \pm  0.11$ & $ 0.25 \pm  0.07$ & $-0.04 \pm  0.28$ & $-0.35 \pm  0.29$ &    16 \\
HSC16apuo       &    3.220 & $-21.94 \pm  0.07$ & $ 5.51 \pm  0.65$ & $ 4.02 \pm  0.38$ & $ 1.62 \pm  0.15$ & $ 3.18 \pm  0.87$ & $-0.01 \pm  0.14$ & $-0.69 \pm  2.55$ & $-0.72 \pm  2.56$ &    17 \\
SNLS06D4eu      &    1.588 & $-21.83 \pm  0.03$ & $10.79 \pm  0.89$ & $ 1.62 \pm  0.07$ & $ 0.65 \pm  0.03$ & $ 2.40 \pm  0.11$ & $ 0.24 \pm  0.04$ & $ 0.36 \pm  0.35$ & $ 0.06 \pm  0.35$ &    18 \\
SNLS07D2bv      &    1.500 & $-20.79 \pm  0.05$ & $18.18 \pm  1.65$ & $ 0.57 \pm  0.05$ & $ 0.67 \pm  0.05$ & $ 1.55 \pm  0.16$ & $ 0.52 \pm  0.06$ & $ 0.28 \pm  0.07$ & $-0.24 \pm  0.09$ &    19 \\
SN2213-1745     &    2.046 & $-21.17 \pm  0.03$ & $39.28 \pm  3.96$ & $ 0.15 \pm  0.03$ & $ 0.10 \pm  0.05$ & $ 0.35 \pm  0.08$ & $ 0.24 \pm  0.04$ & $ 0.02 \pm  0.05$ & $-0.11 \pm  0.07$ &    20 \\
SN2017egm       &    0.030 & $-20.84 \pm  0.03$ & $18.53 \pm  1.48$ & $ 0.78 \pm  0.03$ & $ 0.79 \pm  0.03$ & $ 1.62 \pm  0.16$ & $ 0.32 \pm  0.04$ & $ 0.10 \pm  0.07$ & $-0.21 \pm  0.08$ &    21 \\
SN2015bn        &    0.114 & $-20.77 \pm  0.05$ & $41.32 \pm 12.67$ & $ 0.26 \pm  0.05$ & $ 0.28 \pm  0.05$ & $ 1.05 \pm  0.05$ & $ 0.93 \pm  0.06$ & $ 1.09 \pm  0.09$ & $ 0.03 \pm  0.10$ &    22 \\
\hline
\end{tabular}
\endgroup